\documentclass[aps,twocolumn,prd,showpacs,showkeys,superscriptaddress,nofootinbib]{revtex4-2}

\usepackage{amssymb}

\usepackage{mathrsfs}         %%%%%%%%%%%% This is important. because we used Flower Fonts !!!!!
\usepackage{epsfig}
\usepackage{graphicx}
\usepackage{color}

\usepackage{amssymb}
\usepackage{fixltx2e}
\usepackage{bm}
\usepackage{amsmath,amsfonts,bm,dsfont}
\usepackage{verbatim}
\usepackage{mathrsfs}  %
\usepackage{definitions}
\usepackage{color} \definecolor{darkgreen}{rgb}{0,.5,0}
\usepackage{color}

\newcommand{\rr}[1]{{#1}}
\newcommand{\rrr}[1]{{#1}}
\newcommand{\rv}[1]{{#1}}
\newcommand{\zz}[1]{{#1}}

\def\rA{{\rm A}}
\def\rR{{\rm R}}
\def\rK{{\rm K}}
\def\rM{{\rm M}}
\def\rT{{\rm T}}

\begin{document}
\title{Chiral Magnetic Effect out of equilibrium}
\author{Chitradip Banerjee}
\email{banerjee.chitradip@gmail.com}
\author{Meir Lewkowicz}
\author{Mikhail. A. Zubkov}
\affiliation{Ariel University, Ariel 40700, Israel}

\date{\today}

\begin{abstract}
	We consider relativistic fermionic systems in lattice regularization out of equilibrium. The chiral magnetic conductivity $\sigma_{CME}$ is calculated in spatially infinite system for the case when the chiral chemical potential depends on time while the system initially was in thermal equilibrium at small but nonzero temperature. We find that the frequency dependent $\sigma_{CME}(\omega)$ for any nonzero $\omega$ both in the limits $\omega \ll T$ and $\omega \gg T$ is equal to  its conventional value $1$  when the lattice model approaches continuum limit. Notice that $\sigma_{CME} = 0$ for the case when the chiral chemical potential does not depend on time at all. We therefore confirm that the limit of vanishing $\omega$ is not regular for the spatially infinite systems of massless fermions. 	  
\end{abstract}

\maketitle

\section{Introduction}

It is widely believed that the chiral magnetic effect (CME) \cite{Vilenkin,CME,Kharzeev:2013ffa,Kharzeev:2009pj,SonYamamoto2012} appears out of equilibrium in the presence of external magnetic field and chiral imbalance. The latter may be driven by chiral anomaly due to the parallel magnetic and electric fields \cite{Nielsen:1983rb} or introduced directly by a (time dependent) chiral chemical potential.  Experimental observation of the CME in the first mentioned above case has been reported via measurements of  magnetoresistance of Dirac and Weyl semimetals \cite{ZrTe5}. In the present paper we will discuss the second possibility, i.e. the appearance of electric current in the presence of an external magnetic field driven by the time dependent chiral chemical potential.

It is worth mentioning that the CME belongs to a class of non - dissipative transport effects, which attracted recently attention of many theoreticians and experimentalists both  in condensed matter physics and in high energy physics  \cite{Landsteiner:2012kd,semimetal_effects7,Gorbar:2015wya,Miransky:2015ava,Valgushev:2015pjn,Buividovich:2015ara,Buividovich:2014dha,Buividovich:2013hza}. Several effects of this type were observed  in topological Dirac and Weyl semimetals  \cite{semimetal_effects6,semimetal_effects10,semimetal_effects11,semimetal_effects12,semimetal_effects13,Zyuzin:2012tv,tewary}. Indications of CME have been reported in study of relativistic heavy - ion collisions \cite{Kharzeev:2015znc,Kharzeev:2009mf,Kharzeev:2013ffa}. Lattice simulations suggest appearance of the CME inside vacuum fluctuations \cite{Polikarp}.

 Although the calculation of CME conductivity \cite{Nielsen:1983rb,ZrTe5} obviously requires the use of kinetic theory, the majority of related publications typically refer to other methods. Relatively recently the analysis within the framework of Keldysh technique has been undertaken in \cite{Wu:2016dam}. This technique has been used for  continuum fermion systems in Pauli - Villars regularization. It was argued that in the presence of a time dependent chiral chemical potential the CME effect acquires its conventional expression originally proposed for the equilibrium theory.

It is known presently that this latter  equilibrium version of the CME is actually absent \footnote{In \cite{Valgushev:2015pjn,Buividovich:2015ara,Buividovich:2015ara,Buividovich:2014dha,Buividovich:2013hza} a proof has been given using lattice simulations. In  \cite{Gorbar:2015wya} the question was considered using analytical methods for the specific boundary conditions. In \cite{nogo} Weyl semimetals were considered, where the  absence of equilibrium  CME has been reported. In \cite{nogo2} it was argued that  equilibrium CME contradicts to the Bloch theorem.  In \cite{Z2016} the proof was given based on the representation of the CME conductivity through the momentum space topological invariant. These results were extended to finite temperatures in  \cite{Beneventano:2019qxm}, and to the spatially non - homogeneous interacting systems in \cite{LCZ2021}.}. Therefore, the supposition that it reappears for a time depending chiral chemical potential is intriguing. More specifically, \cite{Wu:2016dam} reports that the electric current along a constant external magnetic field is equal to the standard coefficient $\frac{1}{2 \pi^2}$ multiplied by  magnetic field and chiral chemical potential. The latter depends on frequency, and the frequency is supposed to tend to zero. At the same time it is assumed that the spatial inhomogeneity is taken off before the limit of vanishing frequency is calculated.  In the present paper we analyze this intriguing possibility using the same Keldysh technique as \cite{Wu:2016dam} but for the fermion systems defined in lattice regularization.

Technically we rely on the version of Wigner - Weyl formalism developed for the QFT in \cite{ZW2019,ZZ2019,ZZ2019_} and its unification with Keldysh formalism of quantum kinetic theory. The latter is taken in its path integral form based on \cite{Kamenev}.  The final version of the formalism to be used in the present paper is close to the one of \cite{Shitade, Sugimoto,Sugimoto2006,Sugimoto2007,Sugimoto2008}.

More specifically, we use Wigner transformed Green functions for the calculation of the response of electric current to chiral chemical potential and to external magnetic field. It can be shown using this technique that in equilibrium  the response of electric current (integrated over the system volume) to magnetic field and chiral chemical potential is a topological invariant, including the case of the non - homogeneous systems at finite temperature and in the presence of interactions \cite{LCZ2021}. This topological invariant actually equals to zero identically. Out of equilibrium the mentioned above response looses its topological nature, and, therefore, the CME is back. 

Historically the development of Wigner - Weyl calculus was initiated with the purpose of reformulation of quantum mechanics in the language of functions defined in phase space instead of the language of operators defined in Hilbert space \cite{Weyl_1927}.
The new chapter of mathematics called now "deformation quantization" appeared based on the Wigner - Weyl calculus (for a review  see \cite{Vassilevich_2008,Vassilevich_2015} and references therein). The initial form of Wigner - Weyl formalism refers to the so - called
 Wigner function $W(q,p)$, which is a generalization of quantum mechanical probability distribution \cite{Balazs_1984}. Although Wigner function cannot be treated directly as a probability distribution \cite{Zurek_1991}, it appears to be possible to formulate the fluid analog of quantum entropy flux with the aid of Weyl-Wigner formalism \cite{Bernardini_2017}. Certain quantities of quantum information theory (like von Neumann entropy) have been defined within Wigner - Weyl calculus  \cite{Bernardini_2019,Wlodarz_2003,Bernardini_2017,Bernardini_2}. Practical applications of Wigner - Weyl formalism to quantum mechanics were developed
\cite{Curtright_1998,Bastos_2008,Bernardini_2015,Bernardini_2017}. Besides, Wigner - Weyl calculus has been applied to the anomalous transport, already within the quantum field theory
\cite{Lorce_2011,Lorce_2012,Buot_1990,Buot+Jensen_1990,Miransky+Shovkovy_2015,Prokhorov+Teryaev_2018}.

Keldysh formalism \cite{Keldysh64} has been proposed as the way to construct perturbation theory (similar to that of equilibrium QFT) in the framework of quantum kinetic theory. It has been applied widely both in condensed matter physics and in high energy physics \cite{Bonitz00,BS03,BF06,KB62,Baym62,Schwinger61}. In the limit of thermal equilibrium the Keldysh formalism is naturally reduced to conventional formalism of equilibrium statistical physics \cite{Matsubara55,BdD59,Gaudin60,AGD63}.   Path integral approach to Keldysh technique  \cite{Kamenev} has been developed as an alternative to a more widely used operator formalism \cite{Langreth76,Danielewicz84,CSHY85,RS86,Berges04,HJ98,Rammer07}. The difference between Keldysh formalism and the conventional QFT is the appearance of the so-called Keldysh contour. This is a closed contour in the complex plane of time. In the real time equilibrium QFT the integration occurs only along the real axis of this plane while in the Matsubara formalism the integration is along the imaginary axis. Except for this the formalisms are similar. However, certain silent features are present on the Keldysh side related to the turning points of the Keldysh contour. The naive approach to path integral formulation fails to reproduce the correct expressions for the Green functions even for the simplest non - interacting models. The rigorous lattice regularization is to be used to restore the correct answers \cite{Kamenev,Kamenev2}.   At the same time the operator approach to non-equilibrium diagram technique \cite{LP,Mahan} gives the correct answers immediately without lattice regularization (for the non - interacting stationary systems).

Perturbation expansion of quantum kinetic theory has been applied successfully to investigation of various physical systems  \cite{Bonitz00,BS03,BF06}. The  Schwinger-Dyson equations \cite{Landau56,LW60,Luttinger60} are used within Keldysh technique widely  \cite{Baym62}, and allow to reproduce the  Bogoliubov-Born-Green-Kirkwood-Yvons (BBGKY) sequence of equations \cite{Cercignani88}. Being truncated this sequence gives kinetic equations to be used for the investigation of transport phenomena \cite{CCNS71}, including superconductivity \cite{AG75,LO75,Bonitz00,BS03,BF06,Langreth76,Danielewicz84,CSHY85,RS86,Berges04,HJ98,Rammer07}.
On the high energy physics side the Keldysh formalism was applied to high energy scattering in QCD \cite{CGC} and  relativistic hydrodynamics \cite{hydr}, as well as to various problems in cosmology  \cite{Akh}.

From the very first days of Keldysh technique the notion of Wigner distribution has been used widely in its framework  \cite{Langreth76,Danielewicz84,CSHY85,RS86,Berges04,HJ98,Rammer07,Polkovnikov:2009ys}. In the present paper we use a specific version of Wigner-Weyl formalism developed for quantum field theory (see, for example, \cite{ZZ2019_FeynRule} and references therein). Using this formalism the Hall conductivity has been represented as the topological quantity composed of Wigner-transformed Green functions \cite{Zubkov+Wu_2019}. Besides, it appears to be possible using this formalism to prove that the QHE conductivity is robust to interaction corrections
\cite{Zhang+Zubkov2019,Zhang_2019_JETPL}.
Within Keldysh technique the similar approach has been developed earlier in \cite{Shitade, Sugimoto,Sugimoto2006,Sugimoto2007,Sugimoto2008}. In \cite{Mokrousov,Mokrousov2} the essentially non - homogeneous systems were discussed in this framework.

\section{Basics of Keldysh technique }

\label{SectKeldysh}

%\subsection{Wigner-Weyl formalism in Keldysh technique}

{Let us discuss the quantum field system in the presence of an external magnetic field. The field Hamiltonian is denoted by $\hat\cH$. The average value of a physical quantity represented by an operator  $O[\psi,\bar{\psi}]$ (depending on fermionic fields $\hat{\psi}, \hat{\bar{\psi}}$ taken at time $t$) is given by
	$$
	\langle O \rangle
	=  {\tr} \,\Bigl(\hat{R}(t_i)\, e^{- i \int_{t_i}^{t} \hat\cH dt }  O[\hat{\psi},\hat{\bar{\psi}}] e^{- i \int_{t}^{t_f} \hat\cH dt } e^{ i \int_{t_i}^{t_f} \hat\cH dt }\Bigr).
	$$
	Here $\hat{R}(t_i)$ is density matrix at the initial time moment $t_i<t$. We also fix the final time moment $t_f > t$. In functional integral Keldysh formalism we have (see textbook \cite{Kamenev})
	$$
	\langle O \rangle
	=  \int {\cal D}\bar{\psi} {\cal D} \psi\, O[\psi,\bar{\psi}]
	\exp\left\{\ii \int_C dt \int d^{\rv{D}}x\, \bar{\psi}(t,x) \hat{Q} \psi(t,x) \right\}.
	$$
	By $D$ we denote dimension of space, $\psi$ and $\bar{\psi}$ are the independent Grassmann variables.  For the noninteracting system $\hat{Q}$ is given by $\hat{Q} = i \partial_t-\hat{H}$ with the single particle Hamiltonian $\hat{H}$. Time integration goes along the Keldysh contour $C$. It begins at $t_i$, goes until $t = t_f$, turns back and returns to $t_i$.
Dynamical variables defined on the forward part of the contour $\bar{\psi}_-(t,x)$ and $\psi_-(t,x)$ differ from those of the backward part $\bar{\psi}_+(t,x)$ and $\psi_+(t,x)$.
	
Boundary conditions relate fields defined on the opposite parts of Keldysh contour: $\bar{\psi}_-(t_f,x) =  \bar{\psi}_+(t_f,x)$ and $\psi_-(t_f,x)=\psi_+(t_f,x)$. The integration measure ${\cal D} \bar{\psi} {\cal D} \psi$ includes $\bar{\psi}_+(t_i,x)$, ${\psi}_+(t_i,x)$ and $\bar{\psi}_-(t_i,x)$, ${\psi}_-(t_i,x)$, and contains initial distribution represented by $\hat{R}$:
	\begin{eqnarray}
		&&\langle O \rangle
		=
		\int \frac{{\cal D}\bar{\psi}_\pm {\cal D} \psi_\pm}{{\rm Det}\, (1+\rho)} \, O[\psi_+,\bar{\psi}_+]\nonumber\\
		&&
		\qquad	{\rm exp}\Big( i \int_{t_i}^{t_f} dt \int d^{{D}}x [\bar{\psi}_-(t,x) \hat{Q} \psi_-(t,x)\nonumber\\
		&&-\bar{\psi}_+(t,x) \hat{Q} \psi_+(t,x)]-\int d^{{D}} x\, \bar{\psi}_-(t_i,x) {\rho} \psi_+(t_i,x)\Big) .\label{eq1}
	\end{eqnarray}
	Here $\rho$ is the density operator defined on one particle Hilbert space. 
	\rv{Probability that the one - particle state $|\lambda_i\rangle$ is occupied is given by $\frac{\langle \lambda_i |\rho|\lambda_i\rangle}{1+\langle \lambda_i |\rho|\lambda_i\rangle}$.  }
Let us introduce the Keldysh spinor
	\be
	\Psi = \left(\begin{array}{c}\psi_-\\ \psi_+ \end{array}\right),
	\label{KelPsi}
	\ee
	we represent the average of $O$ as follows
	\begin{eqnarray}
		\langle O \rangle
		&=& \frac{1}{{\rm Det}\, (1+\rho)}\int {\cal D}\bar{\Psi} {\cal D} \Psi \,
		O[\Psi,\bar\Psi]\,\nonumber\\&&
		{\rm exp}\Bigl\{\ii \int_{t_i}^{t_f} dt \int d^{\rv{D}} x \bar{\Psi}(t,x) \hat{\bf Q} \Psi(t,x) \Bigr\} .
	\end{eqnarray}
	% Keldysh action of the model is given by
	%$$
	%S = \int dt d^3x\, \bar{\Psi}(x) \hat{\bf Q} \Psi(x).
	%$$
	 Here
	\begin{eqnarray}
		\hat{\bf Q}
		= \left(\begin{array}{cc}Q_{--} & Q_{-+}\\ Q_{+-} & Q_{++} \end{array} \right).
		\label{KelQ}
	\end{eqnarray}
The correct expressions for the components of this matrix may be obtained either as the continuum limit of the lattice regularized expressions or using operator formalism. The result is
	\begin{eqnarray}
		Q_{++}  &=& -\Big(\ii \partial_t-\hat{H} - \ii \epsilon \frac{1-\rho}{1+\rho}\Big), \quad
		Q_{--}  =  \ii \partial_t-\hat{H} + \ii \epsilon \frac{1-\rho}{1+\rho}, \nonumber \\
		Q_{+-}  &=&  -2\ii \epsilon \frac{1}{1+\rho}, \quad 
		Q_{-+}  = 2\ii  \epsilon \frac{\rho}{1+\rho}
		\label{Qnaive} .
	\end{eqnarray}
\rr{Here $\rho$ is a matrix that gives rise to the initial one - particle distribution $f = \rho (1+\rho)^{-1}$. In case of the distribution depending only on energy (and, in particular for thermal distribution of non - interacting particles)  $\rho = \rho(\hat{H})$ is a function of the one - particle Hamiltonian.} The infinitely small contributions proportional to parameter $\epsilon \to 0$ symbolize the way those functions are understood as the so - called generalized functions (tempered distributions).
For details see Sect. 5.1 of \cite{Kamenev2}.

The Keldysh Green  function $\hat{\bf G}$ is defined as
	\begin{eqnarray}
		&&G_{\alpha_1 \alpha_2}(t,x|t^\prime,x^\prime)
		= \int \frac{{\cal D}\bar{\Psi} {\cal D} \Psi}{\ii{\rm Det}\, (1+\rho)}
		\Psi_{\alpha_1}(t,x) \bar{\Psi}_{\alpha_2}(t^\prime,x^\prime)\nonumber\\&&
		\, \exp\left\{\ii \int_{t_i}^{t_f} dt \int d^{\rv{D}} x\, \bar{\Psi}(t,x) \hat{\bf Q} \Psi(t,x) \right\}.\label{G1}
	\end{eqnarray}
	Here the index $\alpha$ corresponds to components of the Keldysh spinor  \Ref{KelPsi}. The Green function obeys
	$
	\hat{\bf Q}\hat{\bf G}=1 .
	$.
Sometimes a new representation of Keldysh spinors is used that is related to the spinors defined above as follows	
		$$
		\begin{pmatrix}\psi_1 \\ \psi_2 \end{pmatrix}=\frac{1}{\sqrt{2}}\begin{pmatrix}1 & 1 \\ 1 & -1 \end{pmatrix}\begin{pmatrix}\psi_- \\ \psi_+ \end{pmatrix}
		,$$$$
		\begin{pmatrix}\bar{\psi}_1 & \bar{\psi}_2 \end{pmatrix}=\frac{1}{\sqrt{2}}\begin{pmatrix}\bar{\psi}_- & \bar{\psi}_+ \end{pmatrix}\begin{pmatrix}1 & 1 \\ -1 & 1 \end{pmatrix}.
		$$
The Green function in the new representation acquires the triangle form
		\begin{eqnarray}
			\hat{\bf G}^{(K)}&=&-i\langle \begin{pmatrix}\psi_1 \\ \psi_2 \end{pmatrix}\otimes\begin{pmatrix}\bar{\psi}_1 & \bar{\psi}_2 \end{pmatrix}\rangle =\begin{pmatrix}
		G^\rR &G^\rK \\0&G^\rA
	\end{pmatrix}.
			\label{GK}
		\end{eqnarray}
In our paper we will use yet another representation
\bes
	\hat{\bf G}^{(<)}=\begin{pmatrix}
		1&1\\0&1
	\end{pmatrix}\begin{pmatrix}
		G^\rR &G^\rK \\0&G^\rA
	\end{pmatrix}\begin{pmatrix}
		1&-1\\0&1
	\end{pmatrix}
	=	\begin{pmatrix}
		G^\rR &2G^<\\0&G^\rA
	\end{pmatrix} .
	\label{G<}	
\end{eqsplit}
It is related to the Green function defined by Eq. (\ref{G1}) as follows
$
\hat{\bf G}^{(<)} = U \hat{\bf G} V,
$,
where
$
U=\frac{1}{\sqrt{2}}\begin{pmatrix}
	1&1\\0&1
\end{pmatrix}\begin{pmatrix}1 & 1 \\ 1 & -1 \end{pmatrix}=\frac{1}{\sqrt{2}}\begin{pmatrix}2 & 0\\1 &-1 \end{pmatrix}
$
and
$
V=\frac{1}{\sqrt{2}}\begin{pmatrix}
	1&1\\-1&1
\end{pmatrix}\begin{pmatrix}1 & -1 \\ 0 & 1 \end{pmatrix}=\frac{1}{\sqrt{2}}\begin{pmatrix}1 & 0\\-1& 2 \end{pmatrix}
$.
In addition, we have
\begin{eqnarray}
	\hat{\bf Q}^{(<)}&=&V^{-1}\hat{\bf Q} U^{-1}=\begin{pmatrix}Q^\rR  & 2Q^< \\ 0 & Q^\rA  \end{pmatrix}.
\end{eqnarray}
Here we denote
$
Q^\rR =Q^{--}+Q^{-+}$, $Q^\rA =-Q^{-+}-Q^{++}$, $Q^<=-Q^{-+}$. 
As a result
$	G^\rA  = (Q^\rA )^{-1}$,  $G^\rR  = (Q^\rR )^{-1}$, $G^< =-G^\rR  Q^< G^\rA$
with
\bes
G^\rR
&= (\ii \partial_t-\hat{H}e^{+ \epsilon \partial_t})^{-1}
= (\ii \partial_t-\hat{H}+\ii \epsilon )^{-1},
\\
G^\rA  &= ( \ii \partial_t-\hat{H}e^{- \epsilon \partial_t})^{-1}
= ( \ii \partial_t-\hat{H}-\ii \epsilon )^{-1},
\\
G^< &=(G^\rA -G^\rR ) \frac{\rho}{\rho+1}.
\label{Gar_expl}
\end{eqsplit}
The elements of $\hat{\bf Q}^<$ (which is inverse to $\hat{\bf G}^<$) are:
\bes
Q^{<}&=(Q^\rA -Q^\rR )\frac{\rho}{\rho+1} = -2\ii\epsilon \frac{\rho}{\rho+1},
\\
Q^\rR  &= \ii \partial_t-\hat{H}+\ii \epsilon ,	
\\
Q^\rA  &=  \ii \partial_t-\hat{H}-\ii \epsilon .
\label{Qar_expl}
\end{eqsplit}
For more details on the basics of Keldysh technique briefly reviewed above the reader is advised to consult  \cite{Kamenev,Kamenev2}.

 \section{Basics of Wigner - Weyl calculus in Keldysh technique}

 Here we recall basic notions of Wigner - Weyl calculus \cite{Zubkov+Wu_2019,Sugimoto}. In the following the $D+1$ dimensional vectors (with space and time components) are denoted by large Latin letters. For an operator $\hat{A}$ we denote its matrix elements by  $A(X_1,X_2) = \langle X_1 | \hat{A} | X_2 \rangle $. The Weyl symbol of an operator $\hat A$ is then defined as
\be
A_W(X|P)=\int d^{D+1} Y\, e^{\ii Y^\mu P_\mu }A(X+Y/2,X-Y/2)
\label{WignerTr}
\ee
with $ \mu =0,1,...,D$.
 $D+1$ momentum is denoted by $P^\mu=(P^0,p)$, and $P_\mu = (P^0,-p)$. Here  $p$ is spatial momentum with $D$ components.
Below the Weyl symbol of the Keldysh Green function $\hat{\bf G}$ is denoted by $\hat{G}$, while the Weyl symbol of the  Keldysh $\hat{\bf Q}$ is $\hat{Q}$. We omit the subscript $W$ for brevity.
Weyl symbols $\hat{G}$ and $\hat{Q}$ obey the Groenewold equation
\begin{equation}
\hat{Q} * \hat{G} = 1.
\end{equation}
Here the Moyal product $*$ is defined as
\begin{equation}
\left(A* B\right)(X|P) = A(X|P)\,e^{\rv{-}\ii(\overleftarrow{\partial}_{X^{\mu}}\overrightarrow{\partial}_{P_{\mu}}-\overleftarrow{\partial}_{P_{\mu}}\overrightarrow{\partial}_{X^{\mu}})/2}B(X|P).
\end{equation}
In the present paper we consider the situation when electromagnetic potential $A$ corresponds to constant magnetic field and constant spatial components of field strength ${\cal F}^{\mu\nu}$. Expansion in powers of
${\cal F}^{\mu \nu}$ will be used up to the leading order proportional to magnetic field. Introduction of the external gauge potential  results in Peierls substitution $P \to \pi = P- A$. Here $\pi^\mu$ is $D+1$ - dimensional vector similar to $P^\mu$. When the index is lowered its spatial components change sign. The Moyal product may be decomposed as
\begin{equation}
* =  \star~ e^{\rv{-}\ii  \mathcal{F}^{\mu\nu}\overleftarrow{\partial}_{\pi^{\mu}}\overrightarrow{\partial}_{\pi^{\nu}}/2}.
\end{equation}
with
\begin{equation}
\left(A\star B\right)(X|\pi) = A(X|\pi)\,e^{\rv{-\ii(\overleftarrow{\partial}_{X^{\mu}}\overrightarrow{\partial}_{\pi_{\mu}}-\overleftarrow{\partial}_{\pi_{\mu}}\overrightarrow{\partial}_{X^{\mu}})/2}}B(X|\pi).
\end{equation}
 Next, we use expansion of $\hat{Q}$ and $\hat{G}$ in powers of  $\mathcal{F}^{\mu\nu}$ and keep the terms up to the linear one
\begin{equation}
\hat{Q} = \hat{Q}^{(0)}  +\frac{1}{2}\mathcal{F}^{\mu\nu}\hat{Q}_{\mu\nu}^{(1)},\quad
\hat{G} = \hat{G}^{(0)}  +\frac{1}{2}\mathcal{F}^{\mu\nu}\hat{G}_{\mu\nu}^{(1)}.\label{QGK}
\end{equation}
\rr{In the following we omit for simplicity the superscript $^{(0)}$ of the zeroth order contribution to both $G$ and $Q$. For the non - interacting particles with static Hamiltonian and initial distribution $f(\pi_0)$ we have
	\bes
	G^\rR
	&=(\pi_0-\hat{H}(\vec{\pi},x)+\ii \epsilon )^{-1},
	\\
	G^\rA  &= ( \pi_0-\hat{H}(\vec{\pi},x)-\ii \epsilon )^{-1},
	\\
	G^< &=(G^\rA -G^\rR ) f(\pi_0) = 2\pi i \delta(\pi_0-\hat{H}(\vec{\pi}))f(\pi_0).
	\label{Gar_expl}
\end{eqsplit}
The elements of $\hat{\bf Q}^<$ (which is $\star$ -  inverse to $\hat{\bf G}^<$) are:
\bes
Q^{<}&=(Q^\rA -Q^\rR )f(\pi_0) = -2\ii\epsilon f(\pi_0),
\\
Q^\rR  &= \pi_0-\hat{H}(\vec{\pi},x)+\ii \epsilon ,	
\\
Q^\rA  &=  \pi_0-\hat{H}(\vec{\pi},x)-\ii \epsilon .
\end{eqsplit}
}
The Groenewold equation acquires the form
\begin{equation}
\left(\hat{Q}  +\frac{1}{2}\mathcal{F}^{\mu\nu}\hat{Q}_{\mu\nu}^{(1)}\right)\star~ e^{\rv{-}i  \mathcal{F}^{\mu\nu}\overleftarrow{\partial}_{\pi^{\mu}}\overrightarrow{\partial}_{\pi^{\nu}}/2}\left(\hat{G}  +\frac{1}{2}\mathcal{F}^{\mu\nu}\hat{G}_{\mu\nu}^{(1)}\right) = 1.
\label{Groe-F}
\end{equation}
In the zeroth order in $\cal F$ we have $\hat{Q} \star \hat{G}  = 1$, and
$\hat{Q} \star \hat{G}^{(1)}+\hat{Q}^{(1)}\star\hat{G} \rv{-} \ii \hat{Q} \star \overleftarrow{\partial}_{\pi^{\mu}}\overrightarrow{\partial}_{\pi^{\nu}} \hat{G}  = 0$ in the first order. We obtain
\begin{equation}
\hat{G}_{\mu\nu}^{(1)} =-\hat{G} \star  \hat{Q}_{\mu\nu}^{(1)}\star \hat{G}   \rv{-}   \ii\left(\hat{G} \star \partial_{\pi^{\mu}}\hat{Q}  \star\hat{G} \star \partial_{\pi^{\nu}}\hat{Q} \star \hat{G} -(\mu\leftrightarrow \nu)\right)/{2}
.
\label{QGK1}
\end{equation}

Below we will follow closely derivation of  \cite{Sugimoto}. In the non-interacting theory operator of electric current density is given by
$
\hat{j^i}=-\hat{\bar{\psi}} \frac{\partial \hat{Q}}{\partial p_i} \hat{\psi},\quad
i=1,2,\ldots D.
$.
Spatial components of momentum are $p^i=p_i = P^i = - P_i$. The averaged current density as a function of time is given by:
\begin{eqnarray}
\langle j^i(t,x) \rangle
&=&-\frac{\ii}{2}{\tr} \[\hat{\bf G} \hat{{\bf v}}^i\].
\end{eqnarray}
The velocity operator is given by
$
\hat{\bf v}^i = \partial_{p_i}\begin{pmatrix}-Q^{--} & 0 \\ 0 & Q^{++} \end{pmatrix}.
$.
Let us express the velocity operator and current density through the Keldysh Green function written in triangle representation of Eq. (\ref{G<}).
\begin{eqnarray}
\hat{\bf v}_i^{(<)}&=&\partial_{p_i}\frac{1}{2}\begin{pmatrix}2 & 0\\1& 1 \end{pmatrix}\begin{pmatrix}-Q^{--} & 0 \\ 0 & Q^{++} \end{pmatrix}\begin{pmatrix}1 & 0\\1& -2 \end{pmatrix}\nonumber\\&&
=\partial_{p_i}\begin{pmatrix} -Q^{--} & 0 \\ \frac{-Q^{--}+Q^{++}}{2}  & -Q^{++} \end{pmatrix}
\end{eqnarray}
We use that $Q^{--}=Q^\rR +Q^<$, $Q^{-+}=-Q^<$, $Q^{+-}=-Q^\rR +Q^\rA -Q^<$, and $Q^{++}=Q^<-Q^\rA $, and we represent the current density as
\begin{eqnarray}
&&\langle j^i \rangle
=- \frac{\ii}{2}\tr\left[\hat{\bf G}\hat{\bf v}^i\right]
=\frac{\ii}{2}\tr \left(G^\rR \partial_{p_i} Q^\rR -G^\rA \partial_{p_i}Q^\rA \right)\nonumber\\&&
+\frac{\ii}{2}\tr \left(G^\rR \partial_{p_i} Q^<+G^<\partial_{p_i} Q^\rA \right)
+\frac{\ii}{2}\tr \left(G^\rA \partial_{p_i}Q^< +G^<\partial_{p_i} Q^\rR \right)\nonumber
\end{eqnarray}
The second term in this expression is given by $\frac{\ii}{2}\tr\left({\bf G}\partial_{p_i} {\bf Q}\right)^<$.
At the same time the third term is its complex conjugate. We obtain
\be
\langle j^i \rangle
=\frac{\ii}{2}\tr\left(\hat{\bf G}\partial_{p_i}\hat {\bf Q}\right)^\rR
+ \frac{\ii}{2}\tr\left(\hat{\bf G}\partial_{p_i}\hat {\bf Q}\right)^<+{\rm c.c.}
\label{<j>}
\ee
Using Wigner - Weyl calculus we represent the electric current as
\bes
&J^i(X) \equiv \langle j^i(t,x) \rangle
 	= \rv{-} \frac{\ii }{2}\int \frac{d^{\rv{D+1}}\pi}{(2\pi)^{\rv{D+1}}}
\tr\left(\hat{G} (\partial_{\pi_{i}}\hat{Q})\right)^{\rR}\\&
\rv{-} \frac{\ii }{2}\int \frac{d^{\rv{D+1}}\pi}{(2\pi)^{\rv{D+1}}}
\tr\left(\hat{G} (\partial_{\pi_{i}}\hat{Q})\right)^{\rA}		\rv{-}\frac{\ii }{2}\int \frac{d^{\rv{D+1}}\pi}{(2\pi)^{\rv{D+1}}}
\tr\left(\hat{G} (\partial_{\pi_{i}}\hat{Q})\right)^{<}\\&
\rv{-}\frac{\ii }{2}\int \frac{d^{\rv{D+1}}\pi}{(2\pi)^{\rv{D+1}}}
\tr\left((\partial_{\pi_{i}}\hat{Q}) \hat{G}\right)^{<} .\nonumber
\end{eqsplit}
Small imaginary contribution $\pm \ii \epsilon$ in \Ref{Gar_expl} means that the poles of $G^{\rR}$  ($G^{\rA}$) are moved from the real axis of $\omega$ slightly down (up). The integration line may always be closed at infinity. For that we need to use lattice regularization, which adds to our expression factors that suppress expressions standing inside the integral over (complex - valued) $\omega$ at $|\omega| \to \infty$. As a result the first two terms in the above expression vanish.
We obtain
\begin{eqnarray}
J^i(X) &=& \rv{-}
\frac{\ii }{2}\int \frac{d^{\rv{D+1}}\pi}{(2\pi)^{\rv{D+1}}} \tr\left((\partial_{\pi_{i}}\hat{Q})\hat{G}\right)^{<}\nonumber\\&&
\rv{-}\frac{\ii }{2}\int \frac{d^{\rv{D+1}}\pi}{(2\pi)^{\rv{D+1}}}
\tr\left(\hat{G} (\partial_{\pi_{i}}\hat{Q})\right)^{<}.
\label{J Wigner}
\end{eqnarray}
Applying Eqs. (\ref{QGK})-(\ref{QGK1}) we calculate the contribution to the electric current proportional to the external field strength $\mathcal{F}^{\mu\nu}$:
\begin{eqnarray}
{J}^i
&=&    -\frac{1}{4}\int \frac{d^{\rv{D+1}}\pi}{(2\pi)^{\rv{D+1}} }  \tr\Bigl(\hat{G} \star \partial_{\pi^{\mu}}\hat{Q}  \star\hat{G} \star \partial_{\pi^{\nu}}\hat{Q} \star \hat{G}  \partial_{\pi_{i}}\hat{Q} \Bigr)^{<}\mathcal{F}^{\mu\nu}\nonumber\\
&&
-\frac{1}{4}\int \frac{d^{\rv{D+1}}\pi}{(2\pi)^{\rv{D+1}}}  \tr\Bigl(\partial_{\pi_{i}}\hat{Q}  \hat{G} \star \partial_{\pi^{\mu}}\hat{Q}  \star\hat{G} \star \partial_{\pi^{\nu}}\hat{Q} \star \hat{G}  \Bigr)^{<}\mathcal{F}^{\mu\nu}.\nonumber
\end{eqnarray}
The field strength $\cal F$ gives rise to a constant external magnetic field: ${\cal F}_{ij} = - \epsilon_{ijk} B^k$. We express the electric current as
$$
{J}^i = \Sigma^{ijk}  \mathcal{F}_{jk} = - \Sigma^{ijk}\epsilon_{jkl} B_l = \Sigma^{i}_{l,CME} B^l
$$
By $\Sigma^{i}_{l,CME}$ we denote here
\begin{equation}
\frac{\epsilon_{jkl}}{4} \int \frac{d^{\rv{D+1}}\pi}{(2\pi)^{\rv{D+1}} } \tr\left(\partial_{\pi_{i}}\hat{Q}  \left[\hat{G} \star \partial_{\rv{\pi_{j}}}\hat{Q}  \star \partial_{\rv{\pi_{k}}}\hat{G}  \right]\right)^< +{\rm c.c.}\label{MAIN}
\end{equation}
Component of electric current along magnetic field is given by
$
J^i = \Sigma_{CME} B^i
$
with $\Sigma_{CME}$ given by
\begin{equation}
 \frac{\epsilon_{ijk}}{3! 2} \int \frac{d^{\rv{D+1}}\pi}{(2\pi)^{\rv{D+1}} } \tr\left(\partial_{\pi_{i}}\hat{Q}  \left[\hat{G} \star \partial_{\rv{\pi_{j}}}\hat{Q}  \star \partial_{\rv{\pi_{k}}}\hat{G}  \right]\right)^< +{\rm c.c.}\label{MAIN_}
\end{equation}
Now let us discuss briefly the limiting case of an equilibrium system at zero temperature.
In this case the one particle  Hamiltonian $\hat{H}$ does not depend on time. Let us denote by ${\cal G}$ the following expression
$
{\cal G}(x_1,x_2,\omega) \equiv \langle x_1|(\omega - \hat{H})^{-1}|x_2\rangle\label{calG}
$. 
 It has true singularities when $\omega$ tends to one of the energy levels. The time ordered Green  receives the form
$
 G^\rT ( x, x^{\prime}, \omega)
 	= \lim\limits_{\eta\rightarrow 0} {\cal G}(x, x^{\prime}, \omega+\ii\eta \, {\rm sign} \,\omega).
$. 
The retarded Green  function is given by
$
 G^\rR ( x, x^{\prime}, \omega) = \lim\limits_{\eta\rightarrow 0} {\cal G}(x, x^{\prime}, \omega+\ii\eta),\label{GR}
$
and advanced Green is
$
  G^\rA ( x, x^{\prime}, \omega) = \lim\limits_{\eta\rightarrow 0} {\cal G}(x, x^{\prime}, \omega-\ii\eta).\label{GA}
$.
The Matsubara Green's function $G^\rM $ is defined as
$
	G^\rM ( x, x^{\prime},\omega_n) = {\cal G}( x, x^{\prime},\ii\omega),
$
or in terms of imaginary time $\tau$:
$
G^\rM ( x, x^{\prime},\tau)
  = \frac{1}{\beta}\sum\limits_{n=-\infty}^{\infty}e^{-\ii\omega_n \tau}{\cal G}( x, x^{\prime},\ii\omega).\label{GM}
$.
Here the Matsubara frequency $\omega$ is continuous since we discuss zero temperature limit. These relations between the  retarded (advanced)  and Matsubara Green functions may be extended easily also to their Weyl symbols. Then, for example,
$
	G^\rM _W({x},{p},T,\omega) = \int d^{D} y\, e^{-\ii y p}{\cal G}(x+y/2,x-y/2,i\omega )
$.
One can rewrite $\Sigma_{CME}$ defined above in terms of $G^M$ as 
\begin{equation}
  -i \frac{\epsilon_{ijk}}{3! } \int \frac{d^{\rv{D+1}}\pi}{(2\pi)^{\rv{D+1}} } \tr\left(\partial_{\pi_{i}}{Q}_W^M  \left[{G}_W^M \star \partial_{\rv{\pi_{j}}}{Q}_W^M  \star \partial_{\rv{\pi_{k}}}{G}_W^M  \right]\right)\label{MAIN_}
\end{equation}
Here $Q^M$ is inverse to $G^M$:
$
Q^M_W \star G^M_W = 1
$
and
$
Q^M_W = i \omega - H_W \label{QM}
$.
In our previous paper \cite{LCZ2021} we have shown, using the Wigner - Weyl calculus, that in equilibrium the response of $\Sigma_{CME}$ to chiral chemical potential vanishes for a wide range of physical models. Below we will consider corrections due to a time depending chiral chemical potential for the particular lattice regularization.

  \section{Lattice model with time depenging chiral chemical potential}

Here and below we will consider the system regularized using a rectangular lattice. \rr{More specifically, we will consider discretization of spatial coordinates while time remains continuous. However, we will see that the Euclidean version of our model has the structure of lattice regularization with Wilson fermions, in which both imaginary time and spatial coordinates are regularized. It is supposed that external fields do not vary strongly at the distance of the order of lattice spacing. We define our model in such a way that in thermal equilibrium it is reduced to the system with matrix inverse to Matsubara Green function equal to}
$$
  Q^M_W(\pi)=\sum_{\mu=1}^3\gamma^{\mu}g_{\mu}(\pi)-im(\pi)+\gamma^4g_4(\pi_4).
$$
with $\pi_i = P_i-A_i(x)$, $i=1, 2 , 3$ and $\pi_4 = \omega + i A_0(x)$.
  {Here $\gamma^\mu \equiv \gamma_E^\mu = -i \gamma_M^\mu$ ($\mu = 1,2,3$) are Euclidean gamma - matrices expressed through the Minkowski gamma matrices $\gamma_M^\mu$ ($\gamma_E^4 = - \gamma_M^0$):
  $
\gamma^1 = \left(\begin{array}{cccc}
0 & 0 & 0 & -i \\ 0 & 0 & -i & 0\\ 0 & i &  0 & 0\\ i & 0 & 0 & 0 \end{array}\right)   
, \quad \gamma^2 = \left(\begin{array}{cccc}
	0 & 0 & 0 & -1 \\ 0 & 0 & 1 & 0\\ 0 & 1 &  0 & 0\\ -1 & 0 & 0 & 0 \end{array}\right)   
$, $
\gamma^3 = \left(\begin{array}{cccc}
	0 & 0 & -i & 0 \\ 0 & 0 &  & i\\ i & 0 &  0 & 0\\ 0 & -i & 0 & 0 \end{array}\right)   
, \quad \gamma^4 = \left(\begin{array}{cccc}
	0 & 0 & -1 & 0 \\ 0 & 0 & 0 & -1\\ -1 & 0 &  0 & 0\\ 0 & -1 & 0 & 0 \end{array}\right)   
$.}
$g_i=\sin(\pi_i)$ with $i=1,2,3,4$ and $m(\pi)=m^{(0)}+\sum\limits_{i=1}^{4}\left(1-\cos(\pi_i)\right)$. In the massless case we have explicitly $m^{(0)}=0$. \rr{Let us denote also ${\cal Q}(\pi_0,\vec{\pi})$ by
  \begin{eqnarray}
  	 &&Q^M_W(\pi)|_{\pi_4 = - i \pi_0} = \sum_{\mu=1}^3\gamma^{\mu}g_{\mu}(\pi)\label{Qbase}\\&&-i\left(\sum\limits_{i=1}^{3}\left(1-\cos(\pi_i)\right)+\left(1-{\rm ch}(\pi_0)\right)\right)-i \gamma^4 {\rm sh} (\pi_0) \nonumber
  \end{eqnarray}}
In the presence of time depending fields we  cannot use the Matsubara formalism. Instead we use the Keldysh formalism with expressions for $
		\hat{Q}
		= \left(\begin{array}{cc}Q_{--} & Q_{-+}\\ Q_{+-} & Q_{++} \end{array} \right)$
\rr{that in the static case with initial distribution $f(\pi_0) = \rho(\pi_0)/(1+\rho(\pi_0))$ depending only on energy is given by
\begin{eqnarray}
		Q_{++}  &=& -{\cal Q}(\pi_0,\vec{\pi})  + \ii \epsilon \partial_{\pi_0} {\cal Q}(\pi_0,\vec{\pi})  \frac{1-\rho(\pi_0)}{1+\rho(\pi_0)} , \nonumber \\
		Q_{--}  &=&  {\cal Q}(\pi_0,\vec{\pi}) +  + \ii \epsilon \partial_{\pi_0} {\cal Q}(\pi_0,\vec{\pi})  \frac{1-\rho(\pi_0)}{1+\rho(\pi_0)} , \nonumber \\
		Q_{+-}  &=&  -2\ii \epsilon \partial_{\pi_0} {\cal Q}(\pi_0,\vec{\pi}) \frac{1}{1+\rho(\pi_0)} , \nonumber \\
		Q_{-+}  &=& 2\ii  \epsilon \partial_{\pi_0} {\cal Q}(\pi_0,\vec{\pi}) \frac{\rho(\pi_0)}{1+\rho(\pi_0)}
		\label{Qnaive2} .
\end{eqnarray}}
Here $\pi = P - A(X)$.
\rr{The infinitely small terms proportional to $\epsilon$ are chosen in such a way that in the static case the advanced and retarded Green functions are given by the conventional expressions while the lesser Green function is equal to $G^< =(G^\rA -G^\rR ) \frac{\rho(\pi_0)}{\rho(\pi_0)+1}$.}
In the limit of small $\pi_0$ the above expressions give rise to the ones that follow from Eq. (\ref{Qnaive}) after substitution $\bar{\psi}_\pm \to  i \bar{\psi}\gamma^4$. Namely, at $\epsilon \to 0$ we obtain
	\begin{eqnarray}
i\gamma^4	Q_{++}  &=& -\Big(\ii \partial_t-\hat{H} - \ii \epsilon \frac{1-\rho}{1+\rho}\Big), \nonumber \\
i \gamma^4	Q_{--}  & = &  \ii \partial_t-\hat{H} + \ii \epsilon \frac{1-\rho}{1+\rho}, \nonumber \\
i \gamma^4	Q_{+-}  &=&  -2\ii \epsilon \frac{1}{1+\rho}, \nonumber \\ 
i \gamma^4	Q_{-+}  & = & 2\ii  \epsilon \frac{\rho}{1+\rho}
	\label{Qnaive22} 
\end{eqnarray}
with 
$$
\hat{H} = - i \gamma^4 \sum_{\mu=1}^3\gamma^{\mu}g_{\mu}(\pi)+\gamma^4 m(\pi)  
$$
Thus we chose the lattice model in such a way that in its continuum limit the standard expressions for the components of Keldysh Green function are reproduced.  

In order to introduce the time depending chiral chemical potential we shift $\pi_0$ by $\mu_5(t) \gamma^5$ \rr{in the terms that do not contain $\epsilon$. Recall that the latter terms are introduced instead of boundary conditions and, therefore, are not affected by modification of the one - particle Hamiltonian localized in the finite region of time.} This gives
\begin{eqnarray}
		Q_{++}  &=& -\Big(\sum_{\mu=1}^3\gamma^{\mu}g_{\mu}(\pi)-im(\vec{\pi},-i\pi_0-i\mu_5(t) \gamma^5)\nonumber\\&&+\gamma^4 g_4(-i\pi_0-i\mu_5(t) \gamma^5) - \gamma^4 \epsilon \,\rr{ e^{-\pi_0 \gamma^4}}\, \frac{1-\rho(\pi_0)}{1+\rho(\pi_0)} \Big), \nonumber \\
		Q_{--}  &=&  \sum_{\mu=1}^3\gamma^{\mu}g_{\mu}(\pi)-im(\vec{\pi},-i\pi_0-i\mu_5(t) \gamma^5)\nonumber\\&&+\gamma^4g_4(-i\pi_0-i\mu_5(t) \gamma^5) + \gamma^4 \epsilon \, \rr{ e^{-\pi_0 \gamma^4}} \frac{1-\rho(\pi_0)}{1+\rho(\pi_0)} , \nonumber \\
		Q_{+-}  &=&  -2\gamma^4 \epsilon \, \rr{ e^{-\pi_0 \gamma^4}} \frac{1}{1+\rho(\pi_0)} , \nonumber \\
		Q_{-+}  &=& 2\gamma^4  \epsilon \, \rr{ e^{-\pi_0 \gamma^4}} \frac{\rho(\pi_0)}{1+\rho(\pi_0)}
		\label{Qnaive2} .
\end{eqnarray}
with
$
m(\vec{\pi},-i\pi_0-i\mu_5(t)\gamma^5)=m^{(0)}+\sum\limits_{i=1}^{3}\left(1-\cos(\pi_i)\right)+\left(1-\cos(-i\pi_0-i\mu_5(t)\gamma^5)\right)
$.
 \rr{Notice that we do not modify the initial distribution $f(\pi_0) = \rho(\pi_0) (1+\rho(\pi_0))^{-1}$ introducing nonzero $\mu_5$. In the absence of the external magnetic field and chiral chemical potential the given system has one Dirac point. Close to it the dependence of energy on momenta has the form of a Dirac cone. This is the region of the Brillouin zone, where we approach continuum limit. The remaining part of the Brillouin zone is irrelevant at low energies.}

 Here we consider the linear response to the time dependent chiral chemical potential and constant magnetic field. Because of the time dependence of the system we use the Keldysh formulation of matrix Green's functions. \rr{In the lesser representation we have
  $
  \hat{Q}=\hat{Q}^{(0)}+\delta \hat{Q}
  $,
  where
   \begin{eqnarray}
  	\hat{ Q}^{(0)}&=&
  	\begin{pmatrix}Q^{(0)\rR}  & 2Q^{(0)<} \\ 0 & Q^{(0)\rA}  \end{pmatrix},
  \end{eqnarray}
  and
  $
  Q^{(0)\rR} = {\cal Q}(\pi_0 + i\epsilon,\vec{\pi}), \quad Q^{(0)\rA} = {\cal Q}(\pi_0 - i\epsilon,\vec{\pi}), \quad Q^{(0)<} = (Q^{(0)\rA} - Q^{(0)\rR})f(\pi_0)
  $
  while
  $
  \delta\hat{Q}= \delta \mu_5(t)\begin{pmatrix}\frac{\partial {\cal Q}}{\partial \pi_0}  & 0 \\ 0 & \frac{\partial {\cal Q}}{\partial \pi_0}  \end{pmatrix}\gamma^5
  $.}
  In the above we assume the sinusoidal time dependence of $\delta \mu_5(t)=\delta \mu_5^{(0)}\cos{\omega_0 t}$.
  In a general case, we have the expression for the conductivity tensor $\Sigma^{ijk}$, given by
  \begin{eqnarray}
  	&&\Sigma^{ijk}=
  	-\frac{1}{4} \int \frac{d^{D+1}\pi}{(2\pi)^{D+1} } \tr\left(\partial_{\pi_{i}}\hat{Q}
  	\left[
  	\hat{G} \star \partial_{\pi_{[j}}\hat{Q}  \star \hat{G}\star \partial_{\pi_{k]}}\hat{Q} \star  \hat{G}
  	\right]\right)^<\nonumber
  	\\\nonumber &-&
  	\frac{1}{4} \int \frac{d^{D+1}\pi}{(2\pi)^{D+1} } \tr\left(
  	\left[
  	\hat{G} \star \partial_{\pi_{[j}}\hat{Q}  \star \hat{G}\star \partial_{\pi_{k]}}\hat{Q} \star  \hat{G}
  	\right]
  	\partial_{\pi_{i}}\hat{Q} \right)^< .\nonumber
  \end{eqnarray}
  Since $\rm cos$ is an even function, we can write
  \begin{equation}
  	\begin{split}
  		\Sigma^{ijk}
  		=
  		-\frac{1}{8} \int \frac{d^{D+1}\pi}{(2\pi)^{D+1} } \tr\left(\partial_{\pi_{i}}\hat{Q}
  		\left[
  		\hat{G} \star \partial_{\pi_{[j}}\hat{Q}  \star \hat{G}\star \partial_{\pi_{k]}}\hat{Q} \star  \hat{G}
  		\right]\right)^<\\
  		-
  		\frac{1}{8} \int \frac{d^{D+1}\pi}{(2\pi)^{D+1} } \tr\left(
  		\left[
  		\hat{G} \star \partial_{\pi_{[j}}\hat{Q}  \star \hat{G}\star \partial_{\pi_{k]}}\hat{Q} \star  \hat{G}
  		\right]
  		\partial_{\pi_{i}}\hat{Q} \right)^< \\+(\omega_0 \leftrightarrow - \omega_0)
  	\end{split}\nonumber
  \end{equation}
  We can write $\Sigma^{ijk}=\rm I+\rm II+(\omega_0 \leftrightarrow - \omega_0)$ where
  $$
  \rm I=-\frac{1}{8} \int \frac{d^{D+1}\pi}{(2\pi)^{D+1} } \tr\left(\partial_{\pi_{i}}\hat{Q}
  \left[
  \hat{G} \star \partial_{\pi_{[j}}\hat{Q}  \star \hat{G}\star \partial_{\pi_{k]}}\hat{Q} \star  \hat{G}
  \right]\right)^<
  $$
  and
  $$
  \rm II=-\frac{1}{8} \int \frac{d^{D+1}\pi}{(2\pi)^{D+1} } \tr\left(
  \left[
  \hat{G} \star \partial_{\pi_{[j}}\hat{Q}  \star \hat{G}\star \partial_{\pi_{k]}}\hat{Q} \star  \hat{G}
  \right]
  \partial_{\pi_{i}}\hat{Q} \right)^<.
  $$
  while inside up to the terms linear in $\delta\mu_5^{(0)}$ we can write
  \rr{\begin{equation}
  	\begin{split}
  		Q^{R} \to \tilde Q^{R} = {\cal Q}(\pi_0 + i\epsilon,\vec{\pi}) +\delta \mu_5^{(0)}e^{\ii\omega_0 t}\frac{\partial {\cal Q}}{\partial \pi_0}\gamma^5,
  		\\
  		Q^{A} \to \tilde  Q^{A} = {\cal Q}(\pi_0 - i\epsilon,\vec{\pi}) +\delta \mu_5^{(0)}e^{\ii\omega_0 t}\frac{\partial {\cal Q}}{\partial \pi_0}\gamma^5\label{GRGA}
  	\end{split}
  \end{equation}}
  The ``lesser" component is the same
  \begin{equation}
  	Q^< =- 2 \gamma^4 \epsilon f(\pi_{\hat{0}}),
  \end{equation}
  and $Q\star G = 1$.
  For the variation of  conductivity, $\Delta \Sigma=\Sigma(\delta\mu_5^{(0)})-\Sigma(0)$, we have the following contribution of $\rm I$ denoted by $\Delta^{\rm I}\Sigma^{ijk}$:
  \begin{eqnarray}
  	&-&\frac{1}{8} \int \frac{d^D\pi}{(2\pi)^D } \tr\left(\partial_{\pi_{i}}\hat{Q}_0
  	\left[
  	\Delta\hat{G} \star \partial_{\pi_{[j}}\hat{Q}_0  \star \hat{G}_0\star \partial_{\pi_{k]}}\hat{Q}_0 \star  \hat{G}_0
  	\right]\right)^<\nonumber\\\nonumber &-&\frac{1}{8} \int \frac{d^D\pi}{(2\pi)^D } \tr\left(\partial_{\pi_{i}}\hat{Q}_0
  	\left[
  	\hat{G}_0 \star \partial_{\pi_{[j}}\hat{Q}_0  \star \Delta\hat{G}\star \partial_{\pi_{k]}}\hat{Q}_0 \star  \hat{G}_0
  	\right]\right)^<\nonumber\\\nonumber &-&\frac{1}{8} \int \frac{d^D\pi}{(2\pi)^D } \tr\left(\partial_{\pi_{i}}\hat{Q}_0
  	\left[
  	\hat{G}_0 \star \partial_{\pi_{[j}}\hat{Q}_0  \star \hat{G}_0\star \partial_{\pi_{k]}}\hat{Q}_0 \star  \Delta\hat{G}
  	\right]\right)^<,
  \end{eqnarray}
  where $\Delta \hat{G}=-\hat{G}_0\star\Delta \hat{Q} \star \hat{G}_0$, \rr{$\hat{G}_0$ is the Green function with $\mu_5=0$, and
  $$
  \Delta \hat{Q}=\frac{1}{2}\delta \mu_5^{(0)}e^{\ii\omega_0 t}\begin{pmatrix}\frac{\partial {\cal Q}}{\partial \pi_0}  & 0 \\ 0 & \frac{\partial {\cal Q}}{\partial \pi_0}  \end{pmatrix}\gamma^5
  $$
  We also denote by $\hat{Q}_0$ the matrix $\hat{Q}$ with $\mu_5=0$ inserted.}

  \section{Calculation in spatially homogeneous case}

  Let us restrict ourselves to the case with spatial homogeneity. First, we derive a useful identity for the star product containing $e^{\ii\omega_0 t}$: 
 $  	{\rm exp} (\ii\omega_0 t) \star h(\omega)
  	=
  	{\rm exp} (\ii\omega_0 t) e^{-\ii\overleftarrow{\partial_t} \partial_\omega/2} h(\omega) =  {\rm exp} (\ii\omega_0 t) e^{\omega_0 \partial_\omega/2} h(\omega) =  {\rm exp} (\ii\omega_0 t) h(\omega+\omega_0/2)$.
  Therefore,
 \rr{ \begin{eqnarray}
  	\Delta\hat{G}&=&-\hat{G}_0\star \delta \mu_5^{(0)}e^{\ii\omega_0 t}\begin{pmatrix}\frac{\partial {\cal Q}}{\partial \pi_0}  & 0 \\ 0 & \frac{\partial {\cal Q}}{\partial \pi_0}  \end{pmatrix}\gamma^5\star \hat{G}_0\\\nonumber&=&- \delta \mu_5^{(0)}e^{\ii\omega_0 t}\hat{G}_0(\omega-\frac{\omega_0}{2})\begin{pmatrix}\frac{\partial {\cal Q}}{\partial \pi_0}  & 0 \\ 0 & \frac{\partial {\cal Q}}{\partial \pi_0}  \end{pmatrix}\gamma^5\hat{G}_0(\omega+\frac{\omega_0}{2}).
  \end{eqnarray}}	
  Here $\omega = \pi_0$ and we omit for simplicity the dependence of $\hat{G}$ on $\vec{\pi}$.
  Let us define $K^{[\pm]}\equiv K(\omega\pm \omega_0/2),~~K^{[0]}\equiv K(\omega)$. This gives the following expression for $\Delta^{\rm I}\Sigma^{ijk}$:
  \begin{eqnarray}
  	&&\Delta^{\rm I}\Sigma^{ijk}=\frac{\delta \mu_5^{(0)}e^{\ii\omega_0 t}}{8} \int \frac{d^{D+1}\pi}{(2\pi)^{D+1} } \tr\Big(\partial_{\pi_{i}}\hat{Q}^{[0]}_0
  	\Big[
  	\hat{G}^{[-]}_0 \frac{\partial Q^{[0]}_{\rr{diag}}}{\partial \pi_0}\nonumber\\&& \gamma^5 \hat{G}^{[+]}_0\partial_{\pi_{[j}}\hat{Q}^{[+]}_0  \hat{G}^{[+]}_0 \partial_{\pi_{k]}}\hat{Q}^{[+]}_0 \hat{G}^{[+]}_0
  	\Big]\Big)^<\label{Jlesser0} \\\nonumber &+&\frac{ \delta \mu_5^{(0)}e^{\ii\omega_0 t}}{8} \int \frac{d^{D+1}\pi}{(2\pi)^{D+1} } \tr\Big(\partial_{\pi_{i}}\hat{Q}^{[0]}_0
  	\Big[
  	\hat{G}^{[-]}_0\nonumber\\&& \partial_{\pi_{[j}}\hat{Q}^{[-]}_0 \hat{G}^{[-]}_0 \frac{\partial Q^{[0]}_{\rr{diag}}}{\partial \pi_0}\gamma^5\hat{G}^{[+]}_0 \partial_{\pi_{k]}}\hat{Q}^{[+]}_0 \hat{G}^{[+]}_0
  	\Big]\Big)^<\\\nonumber &+&\frac{ \delta \mu_5^{(0)}e^{\ii\omega_0 t}}{8} \int \frac{d^{D+1}\pi}{(2\pi)^{D+1} } \tr\Big(\partial_{\pi_{i}}\hat{Q}^{[0]}_0
  	\Big[\hat{G}^{[-]}_0
  	\partial_{\pi_{[j}}\hat{Q}^{[-]}_0  \hat{G}^{[-]}_0\nonumber\\&& \partial_{\pi_{k]}}\hat{Q}^{[-]}_0 \hat{G}^{[-]}_0 \frac{\partial Q^{[0]}_{\rr{diag}}}{\partial \pi_0}\gamma^5 \hat{G}^{[+]}_0
  	\Big]\Big)^<.\label{Jlesser}
  \end{eqnarray}
  \rr{Here $Q^{[0]}_{\rr{diag}}$ is the diagonal part of Keldysh matrix. The off - diagonal component is absent here because the introduction of $\mu_5$ does not affect the initial distribution. As a result function $f(\pi_0)$ entering $G^<$ in the above expression appears without derivative. In the above expression the second row vanishes identically because it contains the complete derivative over $\pi_i$. The third row may be considered in a way similar to that of the second one. Therefore, let us consider the first row of the last expression.
We insert into it the rows containing $\partial_{\pi_0} (\hat{Q}^{[0,\pm]A}_0-\hat{Q}^{[0,\pm]R}_0)=0$, which are equal to zero identically. If there would be the nondiagonal element in $\partial Q^{[0]}_{\rr{diag}}$, the terms proportional to $df(\pi_0)/d \pi_0$ would appear. As it was mentioned above, these terms are absent because introduction of chiral chemical potential in our model does not affect initial distribution according to our conventions. Physically this corresponds to the time - depending $\mu_5(t)$ that is vanishing at $t \to t_i$, and we consider here the response to the harmonics of this signal with frequency $\omega_0$. One can see, that in the resulting expression most of the terms cancel each other and we are left with
 \begin{eqnarray}
	&&\Delta^{\rm I}\Sigma_1^{ijk}=\frac{\delta \mu_5^{(0)}e^{\ii\omega_0 t}}{8} \int \frac{d^{D+1}\pi}{(2\pi)^{D+1} } \tr\Big(\partial_{\pi_{i}}\hat{Q}^{[0]}_0
	\Big[
	\hat{G}^{[-]}_0 \frac{\partial Q^{[0]}_{\rr{diag}}}{\partial \pi_0}\nonumber\\&& \gamma^5 \hat{G}^{[+]}_0\partial_{\pi_{[j}}\hat{Q}^{[+]}_0  \hat{G}^{[+]}_0 \partial_{\pi_{k]}}\hat{Q}^{[+]}_0 \hat{G}^{[+]}_0
	\Big]\Big)^<\nonumber\\ &&=
\zz{-}	\frac{\delta \mu_5^{(0)}e^{\ii\omega_0 t}}{8}\int_{-\infty + i 0}^{\infty + i0} d\pi_0 \int \frac{d^{D}\vec{\pi}}{(2\pi)^{D+1} }f(\pi_0+\omega_0/2)\nonumber\\&& \tr\Big(\partial_{\pi_{i}}{\cal Q}^{[0]}_0
	\Big[
	{\cal G}^{[-]}_0 \partial_{\pi_0} {\cal Q}_0^{[0]} \gamma^5 {\cal G}^{[+]}_0\partial_{\pi_{[j}}{\cal Q}^{[+]}_0  {\cal G}^{[+]}_0 \partial_{\pi_{k]}}{\cal Q}^{[+]}_0 {\cal G}^{[+]}_0
	\Big]\Big)
	\nonumber\\ &&\zz{+}
	\frac{\delta \mu_5^{(0)}e^{\ii\omega_0 t}}{8} \int_{-\infty - i 0}^{\infty - i0} d\pi_0 \int \frac{d^{D}\vec{\pi}}{(2\pi)^{D+1} }f(\pi_0+\omega_0/2) \nonumber\\&& \tr\Big(\partial_{\pi_{i}}{\cal Q}^{[0]}_0
	\Big[
	{\cal G}^{[-]}_0 \partial_{\pi_0} {\cal Q}_0^{[0]} \gamma^5 {\cal G}^{[+]}_0\partial_{\pi_{[j}}{\cal Q}^{[+]}_0  {\cal G}^{[+]}_0 \partial_{\pi_{k]}}{\cal Q}^{[+]}_0 {\cal G}^{[0]}_0
	\Big]\Big)\nonumber\\ &&\rrr{	+
		\frac{\delta \mu_5^{(0)}e^{\ii\omega_0 t}}{8} \int \frac{d^{D+1}\pi}{(2\pi)^{D+1} }\,\Big(f(\pi_0-\omega_0/2)-f(\pi_0+\omega_0/2)\Big)} \nonumber\\&& \tr\Big(\partial_{\pi_{i}}\hat{Q}^{[0]R}_0
		\Big[
		(\hat{G}^{[-]A}_0-\hat{G}^{[-]R}_0)\partial_{\pi_0} Q^{[0]A} \gamma^5 \hat{G}^{[+]A}_0\nonumber\\&&\partial_{\pi_{[j}}\hat{Q}^{[+]A}_0  \hat{G}^{[+]A}_0 \partial_{\pi_{k]}}\hat{Q}^{[+]A}_0 \hat{G}^{[+]A}_0
		\Big]\Big)\nonumber
\end{eqnarray}
Here by $\cal G$ we denote the Green function that has true singularities at the values of $\pi_0$ coinciding with the energy levels (the analogue of Eq. (\ref{calG})). Advanced and retarted Green functions are expressed through $\cal G$ according to Eqs. (\ref{GRGA}) while the Matsubara Green function is given by $\hat{G}_0$, in which we substitute $\pi_0$ by $i \pi_4$. Besides, ${\cal Q} = {\cal G}^{-1}$. We can add to the above integrals over $\pi_0$ the integrals over $\pi_0 = \pm i \pi$, which cancel each other identically due to the periodicity of the Green function ${\cal G}(\pi_0,\vec{\pi}) = {\cal G}(\pi_0 + 2 \pi i,\vec{\pi})$. At this point we also require that the initial distribution is Fermi distribution with temperature equal to $T = 1/N$ in lattice units ($N$ is integer). As a result $f(\pi_0 + 2\pi i) = f(\pi_0)$.  In addition we add the integrals over $\int_{\infty- i \pi}^{\infty+i\pi}$ and $\int_{-\infty+ i \pi}^{-\infty-i\pi}$. Both of them vanish because of the function of $\pi_0$ in denominator of $\cal G$ that grows exponentially with ${Re}\,\pi_0$. Now in the integration over $\pi_0$ in the last row the integration contour may be closed in the upper half of the complex plane, while in the previous row the integration contour may be closed in the lower half of the complex plane. Fermi distribution has poles at the values $\pi_0 =  i \omega_n$, where $\omega_n$ is Matsubara frequency. We come to 
\begin{eqnarray}
	&&\Delta^{\rm I}\Sigma_1^{ijk}=\frac{\delta \mu_5^{(0)}e^{\ii\omega_0 t}}{8} \int \frac{d^{D+1}\pi}{(2\pi)^{D+1} } \tr\Big(\partial_{\pi_{i}}\hat{Q}^{[0]}_0
	\Big[
	\hat{G}^{[-]}_0 \frac{\partial Q^{[0]}_{\rr{diag}}}{\partial \pi_0}\nonumber\\&& \gamma^5 \hat{G}^{[+]}_0\partial_{\pi_{[j}}\hat{Q}^{[+]}_0  \hat{G}^{[+]}_0 \partial_{\pi_{k]}}\hat{Q}^{[+]}_0 \hat{G}^{[+]}_0
	\Big]\Big)^<\nonumber\\ &&=
	\frac{2\pi T \delta \mu_5^{(0)}e^{\ii\omega_0 t}}{8}\sum_{\pi_4 = \omega_n} \int \frac{d^{D}\vec{\pi}}{(2\pi)^{D+1} } \tr\Big(\partial_{\pi_{i}}\hat{Q}^{[-]M}_0\nonumber\\ &&
	\Big[
	\hat{G}^{[--]M}_0 \partial_{\pi_4} Q^{[-]M} \gamma^5 \hat{G}^{[0]M}_0\partial_{\pi_{[j}}\hat{Q}^{[0]M}_0  \hat{G}^{[0]M}_0 \partial_{\pi_{k]}}\hat{Q}^{[0]M}_0 \hat{G}^{[0]M}_0
	\Big]\Big)\nonumber\\ &&\rrr{	+
		\frac{\delta \mu_5^{(0)}e^{\ii\omega_0 t}}{8} \int \frac{d^{D+1}\pi}{(2\pi)^{D+1} }\,\Big(f(\pi_0-\omega_0/2)-f(\pi_0+\omega_0/2)\Big)} \nonumber\\&& \tr\Big(\partial_{\pi_{i}}\hat{Q}^{[0]R}_0
		\Big[
		(\hat{G}^{[-]A}_0-\hat{G}^{[-]R}_0)\partial_{\pi_0} Q^{[0]A}\nonumber\\ && \gamma^5 \hat{G}^{[+]A}_0\partial_{\pi_{[j}}\hat{Q}^{[+]A}_0  \hat{G}^{[+]A}_0 \partial_{\pi_{k]}}\hat{Q}^{[+]A}_0 \hat{G}^{[+]A}_0
		\Big]\Big)
\end{eqnarray}
 Here we use notations $K^{[--]}\equiv K(\omega - \omega_0)$, $K^{[++]}\equiv K(\omega + \omega_0)$. The sum is over the Matsubara frequencies $\omega_n = 2 \pi T (n+1/2)$ with $n\in Z$ and $\omega_n \in (-\pi, +\pi]$. Due to periodicity we can also calculate this sum for $n = 0,1,..., N-1$ (here inverse temperature in lattice units is equal to $1/T = N$). The similar expressions are valid for the other two rows of Eq. (\ref{Jlesser0}). }
 One can see that for model with Wilson fermions we have
  $$
  Q^{[+]} = - \gamma^5 [Q^{[-]}]^\dagger \gamma^5, \quad G^{[+]} = - \gamma^5 [G^{[-]}]^\dagger \gamma^5
  $$

  Now we can drop the square parenthesis in the above expression of $\Delta^{\rm I}\Sigma^{ijk}$. In order to calculate the electric current along magnetic field we perform antisymmetrization with respect to indexes $i,j,k$. We also calculate the term $\Delta^{I}\Sigma^{ijk}_3$ (corresponding to the third term in Eq. (\ref{Jlesser})) in the way similar to the above calculation of $\Delta^{I}\Sigma^{ijk}_1$. Besides,  
  $$
  \Delta^{\rm II}\Sigma^{ijk} + (\omega_0 \leftrightarrow - \omega_0)=[\Delta^{\rm I}\Sigma^{ijk} + (\omega_0 \leftrightarrow - \omega_0)]^*
  $$
  We are left with
$$
\Delta \Sigma_{CME} = \frac{1}{4 \pi^2} \sigma_{CME}(\omega_0) \delta \mu_5^{(0)}e^{-\ii\omega_0 t} + (\omega_0 \leftrightarrow - \omega_0)
$$
We introduce here complex - valued frequency dependent chiral magnetic conductivity $\sigma_{CME}(\omega_0)$. Notice that the total value of $\Sigma_{CME}$ should remain real. Therefore, $\sigma_{CME}(-\omega_0)=\bar{\sigma}_{CME}(\omega_0)$, and we have
$$
\Delta \Sigma_{CME} = \frac{1}{4 \pi^2} \sigma_{CME}(\omega_0) \delta \mu_5^{(0)}e^{\ii\omega_0 t} + (c.c.) $$$$= \frac{\delta \mu_5^{(0)}}{2 \pi^2}\,{\rm Re} \, \sigma_{CME}(\omega_0) e^{\ii\omega_0 t}
$$
Let us represent the CME conductivity as the sum of the two terms:
\begin{equation}
\sigma_{CME}(\omega_0) = \sigma^{(I)}_{CME}(\omega_0)+ \sigma^{(II)}_{CME}(\omega_0) 
\end{equation}
where
\begin{eqnarray}
&&\sigma^{(I)}_{CME}(\omega_0) = \frac{1}{2}(\tilde{\sigma}^{(I)}_{CME}(\omega_0)+ [\tilde{\sigma}^{(I) }_{CME}(-\omega_0)]^*), \nonumber\\ && \sigma^{(II)}_{CME}(\omega_0) = \frac{1}{2}(\tilde{\sigma}^{(II)}_{CME}(\omega_0)+ [\tilde{\sigma}^{(II) }_{CME}(-\omega_0)]^*)
\end{eqnarray}
Here in the limit of zero temperature the first term may be calculated within Euclidean space - time using Matsubara Green function:
 \begin{eqnarray}
	\tilde\sigma^{(I)}_{CME}(\omega_0)&=&\zz{-}
	\frac{ \epsilon^{ijk} }{3!\rr{2}} \int \frac{d^{D+1}\pi}{(2\pi)^{D-1} }\tr\Big(\gamma^5 \hat{G}^{[0]}_0\partial_{\pi_{j}}\hat{Q}^{[0]}_0  \hat{G}^{[0]}_0 \nonumber\\ &&\partial_{\pi_{k}}\hat{Q}^{[0]}_0 \hat{G}^{[0]}_0
	\partial_{\pi_{i}}\hat{Q}^{[-]}_0
	\hat{G}^{[--]}_0 \frac{\partial Q^{[-]}}{\partial \pi_4}\Big)^M
	\nonumber\\ &\zz{-}&\frac{ \epsilon^{ijk} }{3!\rr{2}} \int \frac{d^{D+1}\pi}{(2\pi)^{D-1} }\tr\Big(\gamma^5\hat{G}^{[++]}_0\partial_{\pi_{i}}\hat{Q}^{[+]}_0
	\hat{G}^{[0]}_0
	\nonumber\\ &&\partial_{\pi_{j}}\hat{Q}^{[0]}_0  \hat{G}^{[0]}_0 \partial_{\pi_{k}}\hat{Q}^{[0]}_0\hat{G}^{[0]}_0\frac{\partial Q^{[+]}}{\partial \pi_4}\Big)^M
\end{eqnarray}
while the second term is to be calculated using the original advanced and retarted Green functions: 
\begin{eqnarray}
	&&\tilde\sigma^{(II)}_{CME}(\omega_0)={-	
		\frac{ \epsilon^{ijk} }{3!\rr{2}}\int_{-\omega_0/2}^{\omega_0/2} d\pi_0 \int \frac{d^{D}\vec{\pi}}{(2\pi)^{D-1} }\,} \nonumber\\&& \tr\Big(\partial_{\pi_{i}}\hat{Q}^{[0]R}_0
		\Big[
		(\hat{G}^{[-]A}_0-\hat{G}^{[-]R}_0)\partial_{\pi_0} Q^{[0]A} \gamma^5 \hat{G}^{[+]A}_0\nonumber\\ &&\partial_{\pi_{j}}\hat{Q}^{[+]A}_0  \hat{G}^{[+]A}_0 \partial_{\pi_{k}}\hat{Q}^{[+]A}_0 \hat{G}^{[+]A}_0
		\Big]\Big)\\\nonumber &&{+	
		\frac{ \epsilon^{ijk} }{3!\rr{2}}\int_{-\omega_0/2}^{\omega_0/2} d\pi_0 \int \frac{d^{D}\vec{\pi}}{(2\pi)^{D-1} }\,} \nonumber\\&& \tr\Big(\partial_{\pi_{i}}\hat{Q}^{[0]R}_0\Big[
		\hat{G}^{[-]R}_0\partial_{\pi_{j}}\hat{Q}^{[-]R}_0  \hat{G}^{[-]R}_0 \nonumber\\ &&\partial_{\pi_{k}}\hat{Q}^{[-]R}_0 \hat{G}^{[-]R}_0\partial_{\pi_0} Q^{[0]R} \gamma^5
		(\hat{G}^{[+]A}_0-\hat{G}^{[+]R}_0)
		\Big]\Big)
\end{eqnarray}
Notice that the expression for $\tilde\sigma^{(I)}_{CME}(\omega_0)$ at $\omega_0 =0 $ is formally divergent. It may be regularized, for example, introducing finite temperature. Then the limit ${\rm lim}_{\omega_0\to 0}\tilde\sigma^{(I)}_{CME}(\omega_0)$ becomes regular. One can see that the above expression standing inside the integral becomes a total derivative with respect to momentum. As a result the integral  vanishes at $\omega_0=0$. For this reason we refer to $\sigma^{(I)}_{CME}$ as to the topological contribution to CME conductivity. The details of the calculation of $\sigma^{(I)}_{CME}(\omega_0)$ at finite temperature are represented in Appendix \ref{AppendixI}. \zz{We notice here two opposite limits. In the limit $T\gg \omega_0$ at $\omega_0 \to 0$ the value of $\sigma^{(I)}_{CME}(\omega_0)$ tends to zero according to the above mentioned analytical results. However, in the opposite limit   $T\ll \omega_0$  the value of $\sigma^{(I)}_{CME}(\omega_0)$ approaches $-1/3$ when the lattice spacing tends to zero. }

Contrary to $\sigma^{(I)}_{CME}(\omega_0)$ the expression for $\sigma^{(II)}_{CME}(\omega_0)$ remains regular at $\omega_0 \to 0$, and we do not need here the regularization by finite temperature. Expression for $\sigma^{(II)}_{CME}(\omega_0)$ at any $\omega_0$ and $T \to 0$ may be calculated directly. We calculate the integral in expression for $\sigma^{(II)}_{CME}(\omega_0)$ using Wolfram Mathematica. In order to calculate $\sigma^{(II)}_{CME}(\omega_0)$ we consider the case $\omega_0 \ll 2 \pi/a$, in which in the integral over momenta we can substitute the lattice Green function by its continuum limit. The corresponding calculations are described in more details in Appendix \ref{AppendixA}. 
We obtain that for $\omega_0 \gg T$ the second contribution to the CME conductivity ${\sigma}^{(II)}_{CME}(\omega_0)$ is equal to $4/3$. At the same time in the opposite limit $\omega_0 \ll T$ we obtain $
\sigma^{(II)}_{CME} \approx 1
$.

\zz{To summarize, both in the limit $T\ll \omega_0$  and in the opposite limit $T\ll \omega_0$ the value of $\sigma_{CME}(\omega_0)$ approaches  conventional value $1$ when the lattice spacing tends to zero. }

 \section{Conclusions}

 In the present paper we consider the system of massless Dirac fermions in the presence of constant external magnetic field and the time depending chiral chemical potential. The lattice regularization is used and response of electric current both to magnetic field and to the chiral chemical potential is calculated. For the direct calculation we use Keldysh technique unified with lattice Wigner - Weyl calculus. The latter is applicable to the lattice systems provided that the inhomogeneity is sufficicntly weak, i.e. variations of external fields at the distance of the order of lattice spacing are negligible. This condition is satisfied always as long as we deal with the lattice regularization of continuum theory.

 We consider the case when the system originally was in thermal equilibrium. In the absence of the dependence of chiral chemical potential on time Keldysh formalism is reduced to Matsubara technique. The latter is defined in Euclidean space - time with imaginary time as a fourth coordinate. For the practical calculations we use the lattice fermion action, which becomes equal to the standard Wilson fermion action after Wick rotation (in the case when the time dependence of chiral chemical potential is off).

 We consider the chiral chemical potential depending on time as $\mu_5 = \mu_5^{(0)} \,{\rm cos}\, \omega_0 t$ and calculate the CME conductivity $\sigma_{CME}$ (i.e. the coefficient in relation $j = \frac{\sigma_{CME}}{2 \pi^2} \mu_5 B$) as a function of frequency $\omega_0$. We separate the obtained expression for $\sigma_{CME}$ into the two contributions. The first one $\sigma^{(I)}_{CME}$ may be calculated using the Matsubara Green functions. This contribution is not well - defined at strictly vanishing temperature for $\omega_0 = 0$. Therefore, we need regularization by finite temperature in order to investigate its behaviour at $\omega_0\to 0$. \zz{ We observe that it tends to zero when $\omega_0$ approaches zero for any finite value of temperature $T \gg \omega_0$. However, in the opposite limit $T\ll \omega_0$ this value approaches to $-1/3$ when the system approaches continuum limit while the ratio $x = \omega_0/T$ is increased.} The second contribution $\sigma^{(II)}_{CME}$  is essentially non - equilibrium. It is expressed through the Advanced/Retarted Green functions. When the lattice system approaches its continuum limit only the small region in momentum space around zero contributes the corresponding expression. In this region of momentum space the continuum limit of the Advanced/Retarded Green functions may be used. \zz{We observe that $\sigma^{(II)}_{CME}$ does not depend on $\omega_0$ and is equal to $4/3$ for $T=0$. At the same time in the limit $T \ll \omega_0$ the first contribution $\sigma^{(I)}_{CME} \to -1/3$. Therefore, we arrive at the conventional value ${1}$ of the CME conductivity in this limit. In the opposite limit $T\gg \omega_0$ we obtain $
 	\sigma^{(II)}_{CME} \approx 1
 	$, and thus the CME conductivity also approaches the conventional value.

 	We illustrate the above mentioned results by Figures \ref{fig.5} and \ref{fig.6}. In Fig. \ref{fig.5} the dependence of the total CME conductivity on $\omega_0$ is represented for three different values of temperature. One can see that for these values of temperature the conductivity approaches its conventional value when $\omega_0$ is decreased. Moreover, there is almost no dependence on temperature for the considered values of $T$. In Fig. \ref{fig.6} we represent the dependence of the CME conductivity on lattice spacing for different values of the ratio $x  = \omega_0/T$ ranged from $x=0.5$ to $x=80$. We also obtain the data on $x=0.1$, the corresponding points on the given plot coincide with those of $x=0.5$. One can see that irrespective of the considered values of $x$ the CME conductivity approaches its conventional value in continuum limit.  } 
 
 \begin{figure}[h]
 	\centering  %
 	\includegraphics[height=5cm]{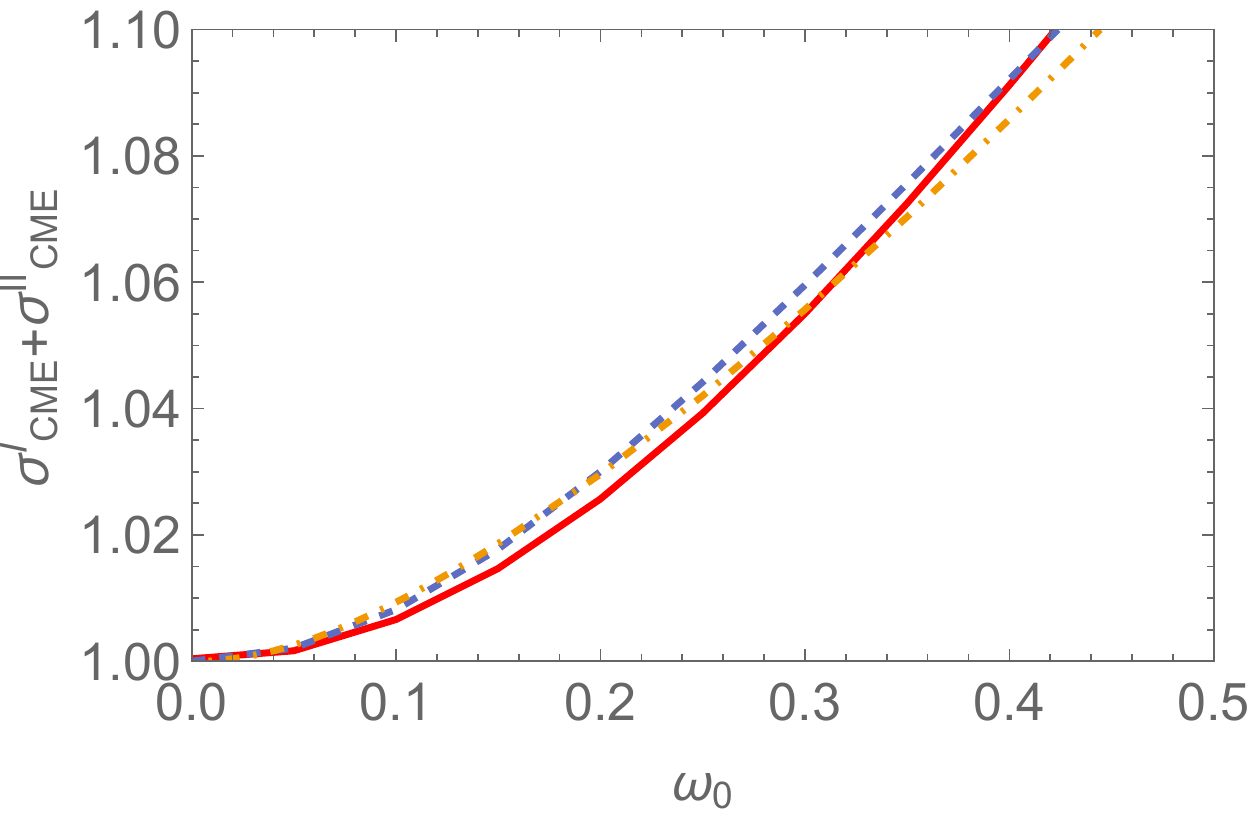} \vspace{1cm} %
 	\caption{We represent here the dependence of  $\sigma^{}_{CME}(\omega_0) $ on $\omega_0$ for the case of the model with Wilson fermions (the imaginary part vanishes). Values of $\omega_0$ are represented in lattice units, i.e. in units of $1/a$, where $a$ is the lattice spacing. The system is considered in the initially equilibrium state with temperatures $T=\frac{1}{10a}$ (solid line), $\frac{1}{20a}$ (dashed line), $\frac{1}{50a}$ (dashed - dotted line). Error bars are of the order of the line widths.}  %
 	\label{fig.5}   %
 \end{figure}
 
 \begin{figure}[h]
 	\centering  %
 	\includegraphics[height=5cm]{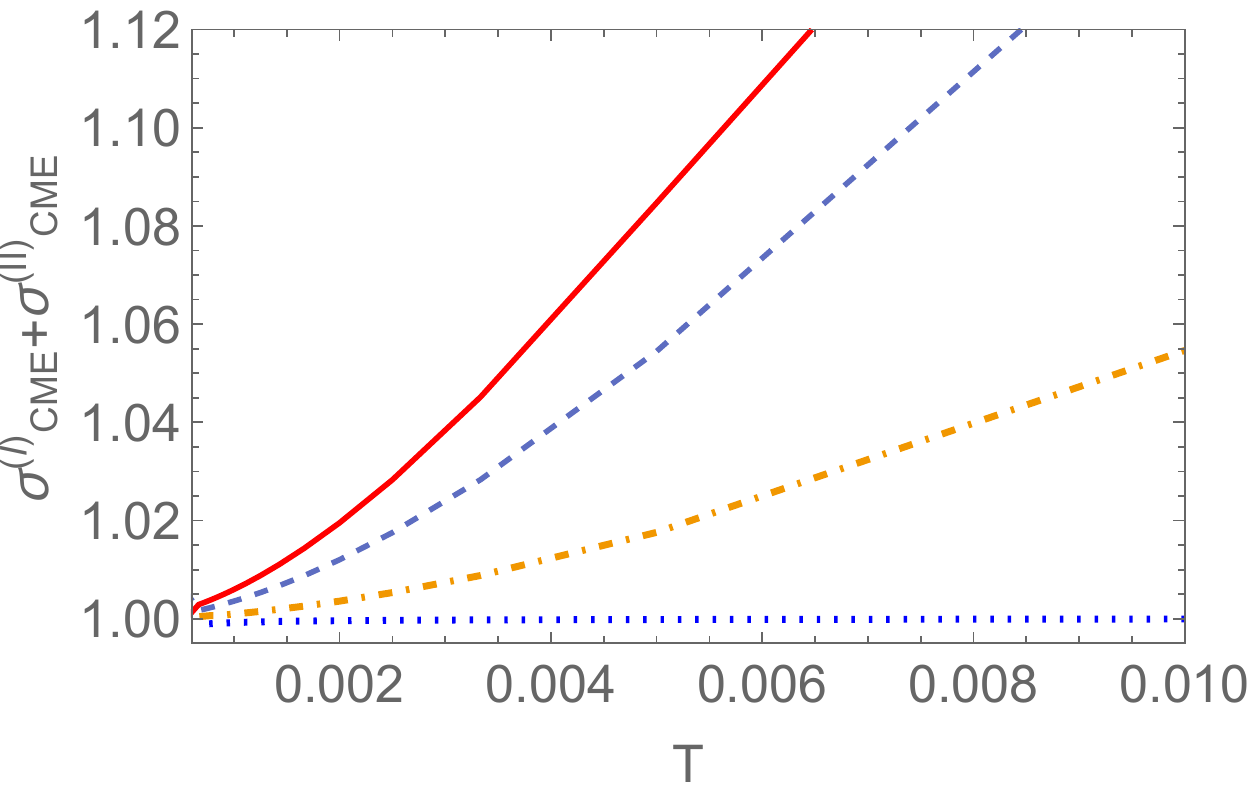} \vspace{1cm} %
 	\caption{We represent here the dependence of  $\sigma^{}_{CME}(\omega_0,T) $ on temperature expressed in lattice units, i.e. on $T a$ for the fixed ratio $x=0.5$ (dotted line), $x = \omega_0/T = 30$ (dashed - dotted  line),  $x = 60$ (dashed line), $x = 80$ (solid  line). The imaginary part of $\sigma^{}_{CME}$ vanishes. Here $a$ is the lattice spacing. One can see that in the continuum limit $a \to 0$ the value of $\sigma^{}_{CME}(\omega_0,T)$ approaches $1$ for all considered values of $x$.  }  %
 	\label{fig.6}   %
 \end{figure}

 \zz{We interpret these numerical results as the presence of the CME at any finite value of $\omega_0 > 0$ and finite temperature $T$. In agreement with the results of \cite{Wu:2016dam} obtained in Pauli - Villars regularization we obtain the conventional value $1$ of the CME conductivity in continuum limit both for $T\ll \omega_0$ and for $T\gg \omega_0$. Besides, we obtain indications that in continuum limit the same value of the CME conductivity is approached irrespective of the values of the ratio $\omega_0/T$ (see Fig. \ref{fig.6}).  
 	
  Recall that in \cite{Wu:2016dam} the similar result was obtained for strictly vanishing temperature in the limit when the spatial non - homogeneity is taken off before the dependence in time of $\mu_5$. In our consideration the system is spatially infinite, which means  the limit, when the spatial non - homogeneity is taken off from the very beginning.}  In the opposite limit  (when spatial inhomogeneity in the chiral chemical potential is taken off first, and the zero frequency limit is taken after this) \cite{Wu:2016dam} predicts vanishing $\sigma_{CME}$ at zero temperature. This is in agreement with our previous result obtained in true equilibrium at finite temperature \cite{LCZ2021}. The important difference of our setup from that of \cite{Wu:2016dam} is that in order to calculate the limit of small $\omega_0$ for the spatially infinite systems we need to consider finite temperature. 
 
 Here an analogy to the physics of graphene is worth to be mentioned. In particular, the value of ordinary electric conductivity in graphene depends strongly on the order of limits used for its calculation. The standard value is obtained using a rather unorthodox procedure when the DC limit $\omega \to 0$ of the AC conductivity is made before the zero disorder strength limit is taken. If the order of limits is reversed, one obtains a different value \cite{Ziegel}.

\zz{To conclude, we observe that, although the  CME effect does not exist in true thermal  equilibrium,  it is back at any nonzero frequency $\omega_0$,  even extremely small, at any nonzero temperature $T$. Provided that $0<T \ll\omega_0$ or $0<\omega_0\ll T$ the conventional value $1$ of the CME conductivity is reproduced in continuum limit. Besides, we obtain indications that the same value of CME conductivity appears in continuum limit for any ratio $\omega_0/T$. 	However, to make the definite conclusion on the CME conductivity for arbitrary ratio $\omega_0/T$  the more detailed numerical analysis is needed, which is out of the scope of the present paper. }

 Our consideration was limited by the non - interacting systems. It would be important to extend it to the interacting ones. We do not exclude at the moment that the interactions will give corrections to the CME conductivity at finite $\omega_0$.   
It would be also interesting to extend the present study to another kind of out of equilibrium CME, i.e. to the appearance of  electric current caused by parallel electric and magnetic fields. We expect certain difficulties in the direct application of Keldysh/Wigner - Weyl techniques to this case. In particular, electron - hole annihilation, and dissipation are to be taken into account. The corresponding study is postponed to future publications.

\appendix

\section{Calculation of $\sigma^{(I)}_{CME}(\omega_0)$ at finite temperature}
\label{AppendixI}

\begin{figure}[h]
	\centering  %
	\includegraphics[height=5cm]{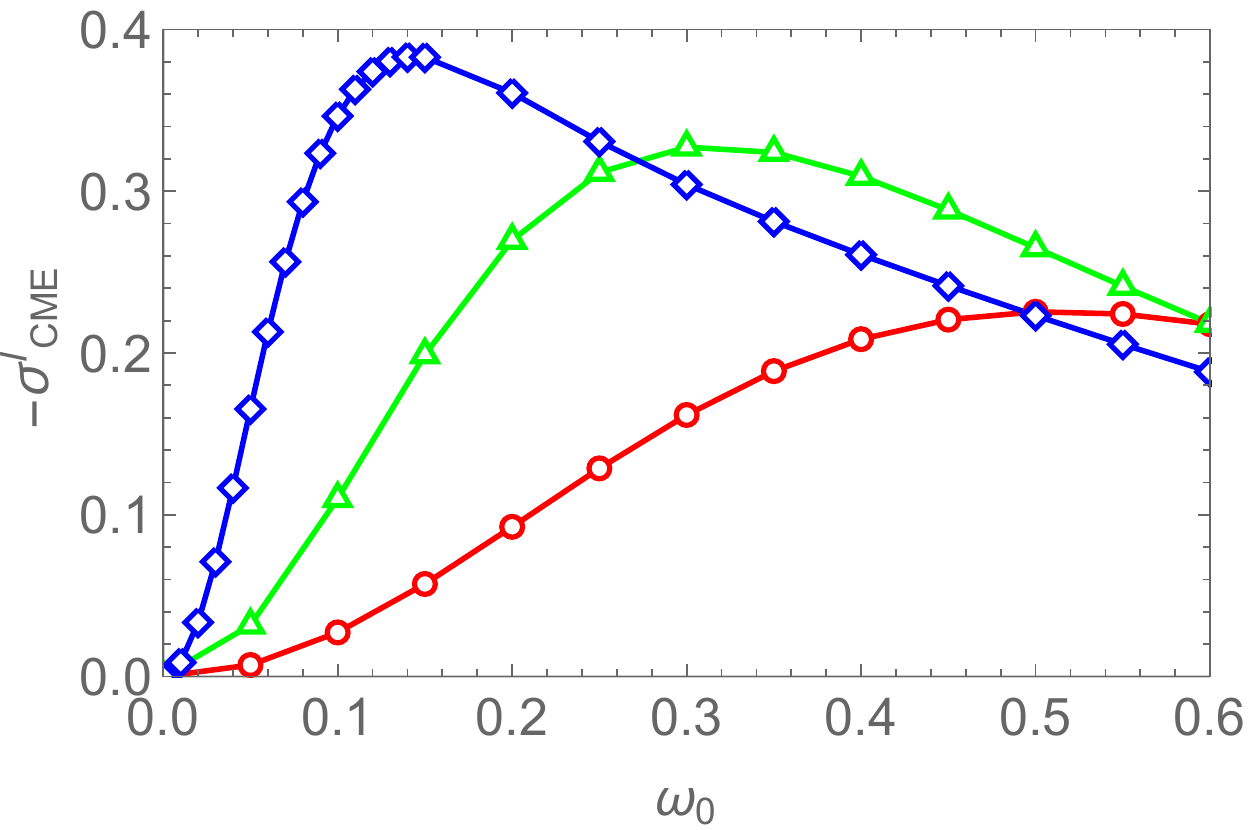} \vspace{1cm} %
	\caption{We represent here the dependence of  $-\sigma^{(I)}_{CME}(\omega_0) $ on $\omega_0$ for the case of the model with Wilson fermions (the imaginary part vanishes). Values of $\omega_0$ are represented in lattice units, i.e. in units of $1/a$, where $a$ is the lattice spacing. The system is considered in the initially equilibrium state with temperatures $T=\frac{1}{10a}$ (circles), $\frac{1}{20a}$ (triangles), $\frac{1}{50a}$ (squares). Error bars are of the order of the sizes of these symbols.}  %
	\label{fig.1}   %
\end{figure}

Explicitly, we have at finite temperature the following expression for $\tilde\sigma^{(I)}_{CME}(\omega_0)$: 
\begin{eqnarray}
	&&\tilde\sigma^{(I)}_{CME}=\zz{-}
	\frac{ 2\pi T \epsilon^{ijk} }{48 \pi^2}\sum_{\pi_4=2 \pi T (n+1/2)} \int d^{3}\pi \tr\Big(\gamma^5 \hat{G}^{[0]}_0\nonumber\\&&\partial_{\pi_{j}}\hat{Q}^{[0]}_0  \hat{G}^{[0]}_0 \partial_{\pi_{k}}\hat{Q}^{[0]}_0 \hat{G}^{[0]}_0
	\partial_{\pi_{i}}\hat{Q}^{[-]}_0
	\hat{G}^{[--]}_0 \frac{\partial Q^{[-]}}{\partial \pi_4}\Big)^M
	\nonumber\\ &&\zz{-}\frac{2\pi T \epsilon^{ijk} }{48 \pi^2} \sum_{\pi_4=2\pi T (n+1/2)} \int d^{3}\pi \tr\Big(\gamma^5\hat{G}^{[++]}_0\partial_{\pi_{i}}\hat{Q}^{[+]}_0
	\hat{G}^{[0]}_0
	\nonumber\\&&\partial_{\pi_{j}}\hat{Q}^{[0]}_0  \hat{G}^{[0]}_0 \partial_{\pi_{k}}\hat{Q}^{[0]}_0\hat{G}^{[0]}_0\frac{\partial Q^{[+]}}{\partial \pi_4}\Big)^M
	\label{finiteT0}
\end{eqnarray}
Here $T = \frac{1}{N a}$, where $a$ is the lattice spacing (in lattice units it is equal to $1$), while $N$ is the number of lattice points in imaginary time direction. Sum over $n$ is  for $n = 0,1,...,N-1$.

This expression does not contain singularities at finite $T$.
One can see that for $\omega_0 = 0$ it vanishes identically because the integral contains the complete derivative with respect to momentum. 
We calculate numerically the dependence of $\sigma_{CME}$ on $\omega_0$. In Fig. \ref{fig.1} we represent the dependence of $\sigma^{(I)}_{CME}$ on $\omega_0$ for the model, which was initially in thermal equilibrium with temperatures $T = \frac{1}{10 a}, \frac{1}{20a}, \frac{1}{50a}$, where $a$ is the lattice spacing (we adopt equal lattice spacings in spatial and imaginary time directions). $\omega_0$ is represented in the units of $1/a$.

\begin{figure}[h]
	\centering  %
	\includegraphics[height=5cm]{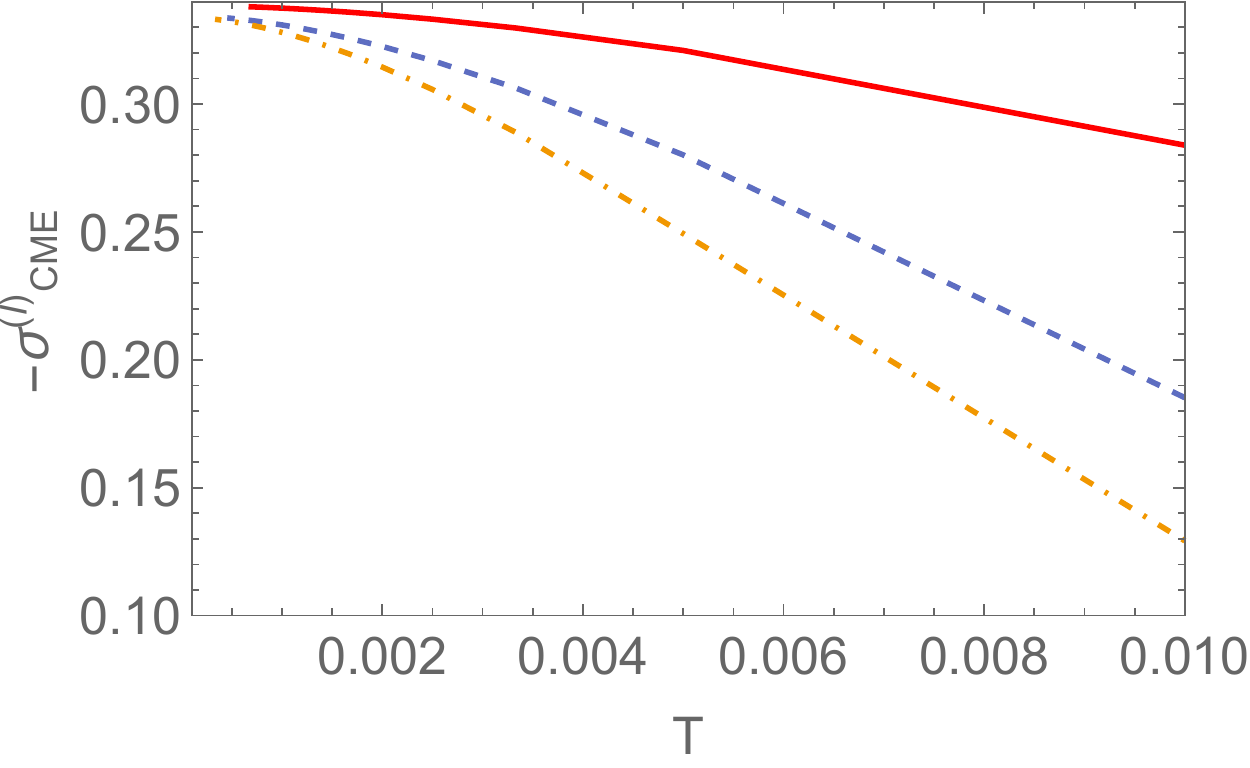} \vspace{1cm} %
	\caption{We represent here the dependence of  $-\sigma^{(I)}_{CME}(\omega_0,T) $ on temperature expressed in lattice units, i.e. on  $T a$, for the fixed ratio $x = \omega_0/T = 30$ (solid line),  $x = 60$ (dashed line), $x = 80$ (dashed dotted  line). The imaginary part of $\sigma^{(I)}_{CME}$ vanishes. Here $a$ is the lattice spacing. One can see that in the continuum limit $a \to 0$ the value of $\sigma^{(I)}_{CME}(\omega_0,T)$ approaches $-1/3$ when the value of $x$ is increased.  }  %
	\label{fig.2}   %
\end{figure}

For $\omega_0 \gg 1/a$ and $T \gg 1/a$ the system should "forget" about discretization, and we would deal effectively with the continuum theory. In this theory there are two dimensional parameters $T$ and $\omega_0$. Any dimensionless quantity is a function of their ratio. In particular, we have
$\sigma^{(I)}_{CME} = f(\omega/T)$. To observe this dependence we should take the case of small $Ta$. First of all, one can see from Fig. \ref{fig.1}  that for $T \gg \omega_0$ the value of $\sigma^{(I)}_{CME}$ remains close to zero. It grows, when $\omega_0$ is increased. 

\zz{We observe that for the fixed value of the ratio $\omega_0/T = x$ the value of $\sigma^{(I)}_{CME}$ depends on $T_{lattice} = 1/N = T_{physical} a_{physical}$, i.e. depends on the lattice spacing $a_{physical}$ expressed in physical units (when we fix the value of $T$ in physical units). Extrapolation to $a_{physical} \to 0$ gives values of  $\sigma^{(I)}_{CME}$ that depend on the ratio $\omega_0/T = x$. In Fig. \ref{fig.2} we represent the data for $x = 30,60,80$. Our numerical results demonstrate that ${\rm lim}_{N \to \infty, x=\omega_0/T}\sigma^{(I)}_{CME}(x)$ approaches $-1/3$ when $x$ grows. }

\section{Calculation of $\sigma^{(II)}_{CME}(\omega_0)$}
\label{AppendixA}

\begin{figure}[h]
	\centering  %
	\includegraphics[height=5cm]{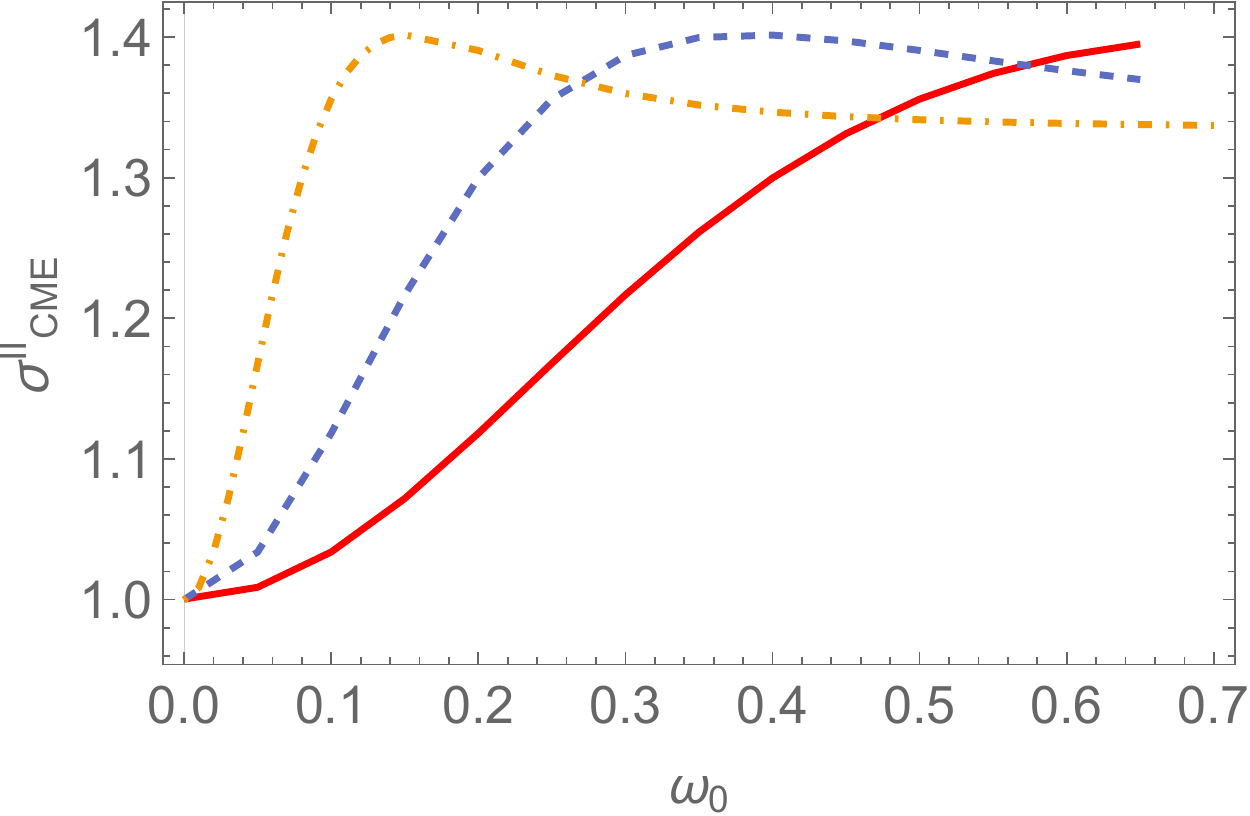} \vspace{1cm} %
	\caption{We represent here the dependence of  $\sigma^{(II)}_{CME}(\omega_0) $ on $\omega_0$ for the case of the model with Wilson fermions (the imaginary part vanishes). Values of $\omega_0$ are represented in lattice units, i.e. in units of $1/a$, where $a$ is the lattice spacing. The system is considered in the initially equilibrium state with temperatures $T=\frac{1}{10a}$ (solid line), $\frac{1}{20a}$ (dashed line), $\frac{1}{50a}$ (dashed - dotted line). For these values of temperature the system effectively approaches continuum limit. Error bars are of the order of the line widths.}  %
	\label{fig.3}   %
\end{figure}

\begin{figure}[h]
	\centering  %
	\includegraphics[height=5cm]{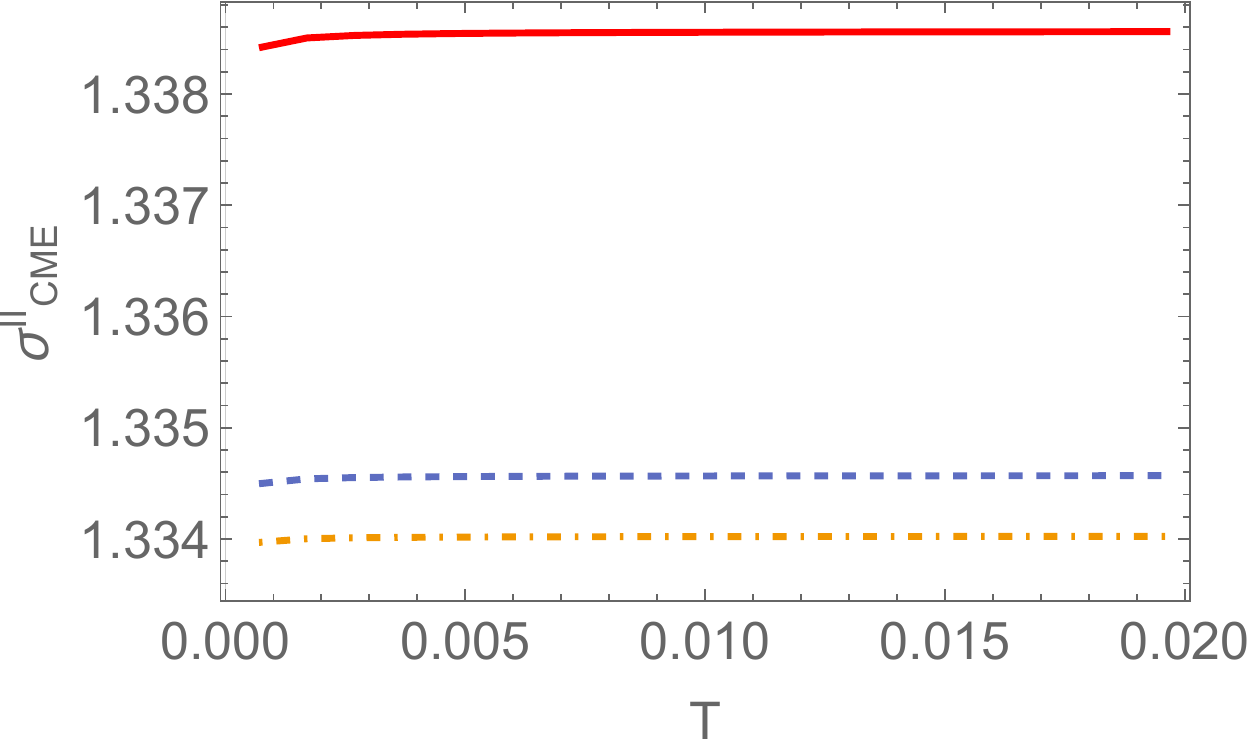} \vspace{1cm} %
	\caption{We represent here the dependence of  $\sigma^{(II)}_{CME}(\omega_0,T) $ on  temperature expressed in lattice units, i.e. on  $T a$, for the fixed ratio $x = \omega_0/T = 30$ (solid line),  $x = 60$ (dashed line), $x = 80$ (dashed dotted  line). The imaginary part of $\sigma^{(II)}_{CME}$ vanishes. Here $a$ is the lattice spacing. One can see that in the continuum limit $a \to 0$ the value of $\sigma^{(II)}_{CME}(\omega_0,T)$ approaches $4/3$ when the value of $x$ is increased.  }  %
	\label{fig.4}   %
\end{figure}

{Let us represent $\sigma^{(II)}_{CME}(\omega_0) = I_1 + I_2$ with 
\begin{eqnarray}
	{\rm I}_1&=&  {	-		\frac{ \epsilon^{ijk} }{3!\rr{2}}\int d\pi_0 \Big(f(\pi_0-\omega_0/2)-f(\pi_0+\omega_0/2)\Big)\int  \frac{d^{D}\vec{\pi}}{(2\pi)^{D-1} }\,} \nonumber\\&& \tr\Big(\partial_{\pi_{i}}\hat{Q}^{[0]R}_0
		\Big[
		(\hat{G}^{[-]A}_0-\hat{G}^{[-]R}_0)\partial_{\pi_0} Q^{[0]A} \gamma^5 \hat{G}^{[+]A}_0\nonumber\\&&\partial_{\pi_{j}}\hat{Q}^{[+]A}_0  \hat{G}^{[+]A}_0 \partial_{\pi_{k}}\hat{Q}^{[+]A}_0 \hat{G}^{[+]A}_0
		\Big]\Big) 
\end{eqnarray}
and
\begin{eqnarray}
&&	{\rm I}_2= 	+
		\frac{ \epsilon^{ijk} }{3!\rr{2}}\int d\pi_0 \Big(f(\pi_0-\omega_0/2)-f(\pi_0+\omega_0/2)\Big)\nonumber\\&&\int \frac{d^{D}\vec{\pi}}{(2\pi)^{D-1} }\,  \tr\Big(\partial_{\pi_{i}}\hat{Q}^{[0]R}_0\Big[
		\hat{G}^{[-]R}_0\partial_{\pi_{j}}\hat{Q}^{[-]R}_0  \hat{G}^{[-]R}_0 \partial_{\pi_{k}}\nonumber\\&&\hat{Q}^{[-]R}_0 \hat{G}^{[-]R}_0  \partial_{\pi_0} Q^{[0]R} \gamma^5
		(\hat{G}^{[+]A}_0-\hat{G}^{[+]R}_0)
		\Big]\Big) 
\end{eqnarray}
Here
\begin{eqnarray}
&&	G^{[-]A}_0-G^{[-]R}_0=2\pi i(\partial_{\pi_0}Q_0)^{-1}\delta(Q_0(\partial_{\pi_0}Q_0)^{-1}).
\end{eqnarray}
After the transformation $Q=i\gamma^4\tilde{Q}$ and $G=-\tilde{G}i\gamma^4$ we have $(\partial_{\pi_0}\tilde{Q}_0)^{-1}=1$ (in the limit of $\pi_0\to 0$ and $\omega_0\to 0$) and consequently we have
$
\tilde{G}^{[-]A}_0-\tilde{G}^{[-]R}_0=2\pi i \delta(\tilde{Q}_0^{[-]})
$.
\begin{eqnarray}
	&&Q^{[-]}=\sum_{\mu=1}^3\gamma^{\mu}g_{\mu}(\vec{\pi})-im(\pi_0-\omega_0/2,\vec{\pi})-i\gamma^4{g_0}(\pi_0-\omega_0/2)\nonumber\\&&=\sum_{\mu=1}^4\gamma^{\mu}\tilde{g}_{\mu}={-i\gamma^4\Big[\sum_{\mu=1}^4\gamma^4\gamma^{\mu}i\tilde{g}_{\mu}+\gamma^4\tilde{m}\Big]=-i\gamma^4\tilde{Q}^{[-]}},
\end{eqnarray}
where {$-i g_0(\pi_0) = g_4 (-i\pi_0)$,  $\tilde{g}_0=-ig_0(\pi_0-\omega_0/2)$}, $\tilde{g}_k=g_k$, and $\tilde{m}=m(\pi_0-\omega_0/2,\vec{\pi})$.
We have {
$$
\tilde{Q}^{[-]}=-\sum_{k=1}^3\gamma^0\gamma^{k}_Mg_{k}(\vec{\pi})-\gamma^0m(\pi_0-\omega_0/2,\vec{\pi})+g_0(\pi_0-\omega_0/2).
$$
Here $-i\gamma^{\mu}=-\gamma^{\mu}_M$ for $\mu = 1,2,3$ and $\gamma^0 =  -\gamma^4$.}
We write $\rm I_1$ as
\begin{eqnarray}
&&	{\rm I}_1=  	-
		2\pi i\frac{ \epsilon^{ijk} }{3!\rr{2}}\int d\pi_0 \Big(f(\pi_0-\omega_0/2)-f(\pi_0+\omega_0/2)\Big)\nonumber\\&& \int \frac{d^{D}\vec{\pi}}{(2\pi)^{D-1} }\, \tr\Big(\delta(\tilde{Q}_0^{[-]})
		\partial_{\pi_0} \tilde{Q}_0^{[0]A} \gamma^5 \tilde{G}^{[+]A}_0\partial_{\pi_{j}}\tilde{Q}^{[+]A}_0 \nonumber\\&& \tilde{G}^{[+]A}_0 \partial_{\pi_{k}}\tilde{Q}^{[+]A}_0\tilde{G}^{[+]A}_0\partial_{\pi_{i}}\tilde{Q}^{[0]R}_0
		\Big)  
\end{eqnarray}
We represent the eigen basis for $\tilde{Q}^{[-]}_0$ as
$
\tilde{Q}^{[-]}_0|\vec{\pi},\pi_0,-,n\rangle=q^{(-)}_n(\vec{\pi},\pi_0)|\vec{\pi},\pi_0,-,n\rangle,
$
with $\langle\vec{\pi},\pi_0,-,n|\vec{\pi},\pi_0,-,n\rangle = 1$ for n=1, 2, 3, 4.
\begin{eqnarray}
	&& \rm I_1=  	-
		2\pi i\frac{ \epsilon^{ijk} }{3!\rr{2}}\int d\pi_0 \Big(f(\pi_0-\omega_0/2)-f(\pi_0+\omega_0/2)\Big)\nonumber\\&&\int \frac{d^{D}\vec{\pi}}{(2\pi)^{D-1} }\,  \sum_n\delta\Big(q^{(-)}_n(\vec{\pi},\pi_0)\Big)\langle\vec{\pi},\pi_0,-,n|
		\partial_{\pi_0} \tilde{Q}_0^{[0]A} \gamma^5\nonumber\\&& \tilde{G}^{[+]A}_0\partial_{\pi_{j}}\tilde{Q}^{[+]A}_0  \tilde{G}^{[+]A}_0 \partial_{\pi_{k}}\tilde{Q}^{[+]A}_0\tilde{G}^{[+]A}_0\partial_{\pi_{i}}\tilde{Q}^{[0]R}_0
	|\vec{\pi},\pi_0,-,n\rangle \nonumber
\end{eqnarray}
where $q^{(-)}_n(\vec{\pi},\pi_0) = \pm \sqrt{\sum_{k=1}^3\tilde{g}_k^2+m^2}+g_4(\pi_0-\omega_0/2)$ and 
$
\Big({-}\sum_{k=1}^3\gamma^0{\gamma}^{k}_M\tilde{g}_{k}(\vec{\pi}){-}\gamma^0m(\pi_0-\omega_0/2,\vec{\pi})\Big)|\vec{\pi},\pi_0,-,n\rangle =\pm\Big( \sqrt{\sum_{k=1}^3\tilde{g}_k^2+m^2}\Big) |\vec{\pi},\pi_0,-,n\rangle.  $
We then write the last expression as
\begin{eqnarray}
	&& \rm I_1=  	-
		2\pi i\frac{ \epsilon^{ijk} }{3!\rr{2}}\sum_n\frac{1}{\partial_{\pi_0}q^{(-)}_n(\vec{\pi},\pi_0)}\int \frac{d^3\vec{\pi}}{(2\pi)^2 }\,\nonumber\\&&\Big(f(\pi_0-\omega_0/2)-f(\pi_0+\omega_0/2)\Big) \nonumber\\&& \langle\vec{\pi},\pi_0,-,n|
		\partial_{\pi_0} \tilde{Q}_0^{[0]A} \gamma^5 \tilde{G}^{[+]A}_0\partial_{\pi_{j}}\tilde{Q}^{[+]A}_0  \tilde{G}^{[+]A}_0 \partial_{\pi_{k}}\tilde{Q}^{[+]A}_0\nonumber\\&&\tilde{G}^{[+]A}_0\partial_{\pi_{i}}\tilde{Q}^{[0]R}_0
	}|\vec{\pi},\pi_0,-,n\rangle|_{\pi_0=\mathcal{E}_n(\vec{\pi}),
\end{eqnarray}
where $\mathcal{E}_n(\vec{\pi})$ is given by solution of equation $q^{(-)}_n(\vec{\pi},\mathcal{E}_n(\vec{\pi}))=0$. We approach continuum limit when $T$ in lattice units approaches zero. In this limit we substitute the Green function by its continuum limit. 
Then expression $q^{(-)}_n(\vec{\pi},\pi_0)=\pm |\vec{\pi}|+(\pi_0-\omega_0/2)$ gives $\partial_{\pi_0}q^{(-)}_n(\vec{\pi},\pi_0) = 1$.
Now
$\partial_{\pi_0}\tilde{Q}_0^{[0]A}\approx 1$, {$\partial_{\pi_i}\tilde{Q}^{[+]A}_0\approx -\gamma^0\tilde{\gamma}^i_M$,  $\partial_{\pi_j}\tilde{Q}^{[+]A}_0\approx -\gamma^0\tilde{\gamma}^j_M$, and $\partial_{\pi_k}\tilde{Q}^{[+]A}_0\approx -\gamma^0\tilde{\gamma}^k_M$.}
We then have
\begin{eqnarray}
&&	\rm I_1= +\frac{ i\epsilon^{ijk} }{24\pi}\sum_n\int d^3\vec{\pi} \Big(f(\pi_0-\omega_0/2)-f(\pi_0+\omega_0/2)\Big)\nonumber\\&& \langle\vec{\pi},\pi_0,-,n|
	\gamma^5 \tilde{G}^{[+]A}_0\gamma^0\tilde{\gamma}^j_M \tilde{G}^{[+]A}_0 \gamma^0\tilde{\gamma}^k_M \tilde{G}^{[+]A}_0\gamma^0\tilde{\gamma}^i_M
	|\vec{\pi},\pi_0,-,n\rangle.\nonumber
\end{eqnarray}
and
\begin{eqnarray}
&&	\rm I_2= -\frac{ i\epsilon^{ijk} }{24\pi}\sum_n\int d^3\vec{\pi} \Big(f(\pi_0-\omega_0/2)-f(\pi_0+\omega_0/2)\Big)\nonumber\\&& \langle\vec{\pi},\pi_0,+,n|\gamma^0\tilde{\gamma}^i_M\tilde{G}^{[-]R}_0
	\gamma^0\tilde{\gamma}^j_M \tilde{G}^{[-]R}_0 \gamma^0\tilde{\gamma}^k_M \tilde{G}^{[-]R}_0\gamma^5 
	|\vec{\pi},\pi_0,+,n\rangle.\nonumber
\end{eqnarray}
Here $\tilde{Q} = -\gamma^0\vec{\pi}\vec{\gamma}_M+\pi_0$. This gives $\tilde{Q}^{[+]} = -\gamma^0\vec{\pi}\vec{\gamma}_M+(\pi_0+\omega_0/2)$, $\tilde{Q}^{[-]} = -\gamma^0\vec{\pi}\vec{\gamma}_M+(\pi_0-\omega_0/2)$, and $\tilde{Q}^{[0]} = -\gamma^0\vec{\pi}\vec{\gamma}_M+\pi_0$.
while  
$
\tilde{G}^{[0]}=-\frac{\gamma^0\vec{\pi}\vec{\gamma}_M+\pi_0}{\vec{\pi}^2-\pi_0^2}$, $\tilde{G}^{[+]}=-\frac{\gamma^0\vec{\pi}\vec{\gamma}_M+(\pi_0+\omega_0/2)}{\vec{\pi}^2-(\pi_0+\omega_0/2)^2}$, $  \tilde{G}^{[-]}=-\frac{\gamma^0\vec{\pi}\vec{\gamma}_M+(\pi_0-\omega_0/2)}{\vec{\pi}^2-(\pi_0-\omega_0/2)^2}. 
$

Direct calculation gives
\begin{eqnarray}
&&	\rm I_1 = I_2=\frac{2}{3} \int_{-\infty}^{+\infty} p^2 d p \frac{  (4 p - 3 \omega_0)}{\omega_0^2 (-2 p + \omega_0-i 0)^2}\nonumber\\&&\Big(\frac{1}{e^{-p\beta}+1}-\frac{1}{e^{(-p+\omega_0)\beta}+1}\Big)\label{I2fin_2}
\end{eqnarray}
We calculate this expression in two opposite cases: when $T \ll \omega_0$ and $T\gg \omega_0$. In the former case we set $T=0$ and then Eq. (\ref{I2fin_2}) is invariant under rescaling $\omega_0 \to \lambda \omega_0$, $\vec{\pi} \to \lambda \vec{\pi}$. As a result both these integrals do not depend on $\omega_0$ for $T = 0$. 
The direct integration gives $\sigma^{(II)}_{CME}=I_1 + I_2 = 4/3$ for $T \ll \omega_0$.

In the opposite limit $T \gg \omega_0$  
 we define $p/T=z$ and obtain
\begin{eqnarray}
&&	\rm I_1 = I_2 \approx \frac{1}{6} \int_{0}^{+\infty}  d z  \frac{1 + {\rm cosh}\, z + 2 z\, {\rm sinh} \, z}{ (1 + {\rm cosh}\, z)^2} = 1/2\nonumber
\end{eqnarray} 
As a result $
\sigma^{(II)}_{CME} = {\rm Re}\, (I_1 + I_2)\approx 1
$ at $T\gg \omega_0$. Besides, we illustrate the behavior of $\sigma_{CME}^{(II)}$ by Figures \ref{fig.3} and \ref{fig.4}.

\end{document}